\documentclass[letterpaper, 12 pt, draftcls, onecolumn]{ieeetran}

\IEEEoverridecommandlockouts
\usepackage{lipsum}
\usepackage{multicol, blindtext}
\usepackage{stfloats}
\usepackage{graphicx}
 \usepackage{tabularx}
 \usepackage{lscape}
 \usepackage{rotating}
 \usepackage{epstopdf}
 \usepackage{graphicx}
 \usepackage[centertags]{amsmath}
 \usepackage{amsfonts}
 \usepackage{amssymb}
\usepackage{subfigure}
 \usepackage{amsthm}
 \usepackage{epstopdf}
 \usepackage{epsfig}
 \usepackage{graphicx}
 \usepackage[centertags]{amsmath}
 \usepackage{amsfonts}
 \usepackage{amssymb}
 \usepackage{cite}
 \usepackage{multirow}
 \usepackage{amsmath}
 \usepackage{amsthm}
 \usepackage[table]{xcolor}
 \usepackage{float}
 \usepackage{amsfonts,amssymb,amsmath,latexsym}
 \usepackage{acronym}
 \usepackage{enumerate}
 \usepackage{comment}
 \usepackage{epsfig}
 \usepackage{subfigure}
 \usepackage{etoolbox}

\newtheorem{theorem}{Theorem}

\title{\LARGE \bf
Ensemble Kalman Filters (EnKF) for State Estimation and Prediction of Two-time Scale Nonlinear Systems with Application to Gas Turbine Engines*
}
\author{Najmeh Daroogheh$^{1}$,  Nader Meskin$^{2}$ and Khashayar Khorasani$^{1}$
\thanks{$^{1}$N. Daroogheh and K. Khorasani are with the Department  of Electrical and Computer  Engineering, Concordia University, Montreal, Quebec, Canada
        {\tt\small n\_daroog@encs.concordia.ca and kash@ece.concordia.ca}}%
\thanks{$^{2}$N. Meskin is with the Department of Electrical Engineering, Qatar University,
        Doha, Qatar
        {\tt\small nader.meskin@qu.edu.qa}}%
}

\renewcommand{\baselinestretch}{1.25}

\begin{document}
\maketitle


\begin{abstract}
In this paper, we propose and develop a methodology for nonlinear systems  health monitoring by modeling the damage and degradation mechanism dynamics as "slow" states that are augmented with the system "fast" dynamical states. This augmentation results in a two-time scale nonlinear system that is utilized for development of health estimation and prediction modules within a health monitoring framework. Towards this end, a two-time scale filtering approach is developed based on the ensemble Kalman filtering (EnKF) approach by taking advantage of the model reduction concept. The performance of our proposed two-time scale ensemble Kalman filters is shown to be superior and less computationally intensive in terms of the equivalent flop (EF) complexity metric when compared to well-known particle filtering (PF) approaches. Our proposed methodology is then applied to a gas turbine engine that is affected by erosion of the turbine as the degradation phenomenon and damage mechanism.
 Extensive comparative studies are conducted to validate and demonstrate the advantages and capabilities of our proposed framework and methodology.
\end{abstract}


\section{INTRODUCTION}\label{I}
Study of systems having a two-time scale separation property is crucial and necessary for development of the next generation of health monitoring and condition-based maintenance (CBM) methods \cite{luo2008, model1}. For example, micro cracks in a spinning shaft, the misalignment of machinery parts during operation, corrosion and erosion processes in the system components, and moisture accumulation in the composite materials of electrical circuits, among others can be modeled as two-time scale systems \cite{luo2008}. Therefore, the focus of this work is to investigate the CBM problem of nonlinear  systems by formulating them as a two-time scale system.

The health monitoring problem involves constructing two main modules, namely the  health estimation  and the propagation prediction modules. While  classical Kalman filters provide a complete and optimal solution to state estimation of linear systems under Gaussian process and measurement noise, the estimation problem for nonlinear systems is still a challenging problem \cite{gillijns2006}. Although, several methods have been proposed to address the estimation and prediction  of nonlinear systems, the  results are either too narrow in applicability or are computationally too involved and challenging \cite{daum1986, krener1996}. Consequently, a number of suboptimal methods have been developed to address practical applications \cite{smith2013}.

The main problem with nonlinear filtering methods that rely on a linearization approximation of the system, as in the extended Kalman filter (EKF), is that they characterize  distribution of states by only first and second moments (the same as in the linear case) and discard higher order moments \cite{gelb1974}. On the other hand, in Monte Carlo-based nonlinear  methods, such as particle filters  (PF) and ensemble Kalman filters (EnKF), one can derive the Fokker-Planck partial differential equations for the time evolution of probability density functions which include all the required information related to prediction error statistics \cite{evensen2003}.

In other words,  EnKF can be considered as extension to classical Kalman filters to  large scale nonlinear systems. EnKF works by propagating an ensemble of $N$ members that capture the mean and covariance of the current state estimate \cite{gillijns2007}. Our main motivation for selecting EnKF as opposed to  PFs has to do with ensuring that our proposed two-time scale filters have capabilities  that can significantly reduce the dimensionality of the resulting overall estimation scheme dynamics as the order of the system increases.

\subsection{Ensemble Kalman Filters (EnKF) Overview}
The ensemble Kalman filters (EnKF)  have been extensively investigated in  estimation and data assimilation methods and applications \cite{triantafyllou2013, ott2004}. EnKF is related to the particle filter (PF) approach where a particle represents an ensemble member. The main difference between these two filters is on the assumption that all the probability distributions involved in the EnKF are assumed to be Gaussian. Under circumstances that this assumption is applicable, the EnKF method is more efficient than  particle filters \cite{evensen2003}.

Ensemble Kalman filters is a Monte Carlo approximation method for Bayesian update problems. There are around one hundred different implementations of the EnKF \cite{evensen2009}. In the EnKF method, the distribution of the system state is represented by selecting a collection of the states, that is designated as an ensemble, and by replacing the covariance matrix by the sample covariance which is computed from the ensembles. Consequently, the probability density function can be advanced in time  by simply advancing each ensemble member \cite{anderson1999}.

The main advantage of the EnKF approach over the classical Kalman filters as well as the extended Kalman filter (EKF) methods is that it {\it{does not}} require {\it{any}} model linearization and can also be used to assimilate asynchronous observations. However, its main disadvantage is considered to be a possible dynamic imbalance and sub-optimality \cite{sakov2008}. The main reason for using EnKF in data assimilation applications is due to the ease of its implementation and the low computational cost and storage requirements \cite{gillijns2006}.

In other Monte Carlo-based approaches, a relatively small set of ensembles are used to estimate  {\it{a priori}} error covariances \cite{anderson1999}. The ensembles are operated in such a manner that they are random samples, however the ensemble members are actually not independent and the EnKF will fuse them appropriately. The advantage of these methods is that advancing of the pdfs in time is achieved by simply advancing each ensemble member individually.

In another group of methods, such as the Kalman square-root filters, the analysis for the {\it{a posteriori}} state update is performed only once for obtaining both the {\it{a posteriori}} state estimation mean and the error covariance matrix. Subsequently, the {\it{a posteriori}} ensemble perturbations (to the mean of the analysis) are generated by transforming the {\it{a priori}} ensemble perturbations to a set of vectors that can represent the {\it{a posteriori}} error covariance matrix. Therefore, the {\it{a posteriori}} analysis is rendered to the subspace of the ensembles. Since there is an infinite set of {\it{a posteriori}} perturbations that can be used to represent the {\it{a posteriori}} error covariance matrix, numerous methods can be applied following the works in \cite{kwiatkowski2015}. An iterative extension to the ensemble Kalman filter has been developed in \cite{lorentzen2011} to improve the estimation capabilities of the filter in case that the relationship between the measurements and the system states is nonlinear.

\subsection{Singular Perturbation and Two-Time Scale Systems Overview}
In this work, our goal is to develop an EnKF estimation framework for two-time scale systems known as singularly perturbed systems. Singularly perturbed systems are known as systems that are quantified by a discontinuous dependence of the system properties on a small perturbation parameter that is commonly denoted by $\epsilon$. Many physical systems, such as electrical power systems, electronic systems, mechanical systems, biological systems, economical systems and Quantum physics are examples of singularly perturbed systems. These systems do exhibit a two-time scale behavior known as the fast and slow dynamics. The two-time scale property makes the analysis and control of these systems and their implementation more challenging  than conventional ``regular" systems \cite{khalil1996nonlinear}.
Study of systems with time-scale separation is now necessary for development of the next generation of condition-based maintenance and failure prediction methods \cite{model1}.

The problem of linear filtering of linear stochastic singularly perturbed systems was first considered in \cite{rauch} in which the main estimation framework is developed for continuous-time systems with a composite type of filters. Exact decomposition of fast and slow states in design of Kalman filters was proposed in \cite{gajic}. The filtering methodology based on fast-slow decomposition of Kalman filter gains has also been addressed in \cite{shen}. Although, most of the work on singularly perturbed systems have been developed for continuous-time systems, discrete-time singularly perturbed systems have also been extensively studied in \cite{naidu_1}.

The nonlinear filtering problem of nonlinear singularly perturbed systems has been investigated in {\it{only}} a few work \cite{udem, pf}. In \cite{udem} sufficient conditions for solvability of the filtering problem in nonlinear singularly perturbed systems is obtained based on an $H_2/H_{\infty}$ approach, nevertheless the approximations to the filter gains were not addressed. In \cite{pf}, the authors have proposed a hybrid homogenized method based on the particle filter approach to approximate the nonlinear system states. This method is computationally very complex and its complexity grows exponentially as the number of states increases, and therefore it is not computationally efficient.

Inspired from a local EnKF method that is proposed in \cite{ott2004}, in which the idea of covariance localization is introduced, we take advantage of this covariance definition to reduce the dimension of the covariance matrix in our estimation scheme. We develop the EnKF in the ``dominant" direction of the state space (i.e., the slow state) which also results in a reduced ensemble size as well. Subsequently, a correction is incorporated in the estimated slow states by taking into account the effects of the fast system states. In what follows, more details on our proposed two-time scale estimation scheme based on the EnKF will be provided.

To summarize, the {\underline{main contributions}} of this work can be stated as follows:
\begin{enumerate}
\item A novel two-time scale modeling strategy is proposed for representing the degradation and gradual damage phenomenon in mechanical industrial systems based on the singular perturbation theory,
\item A novel two-time scale filtering methodology is developed based on the two-time scale ensemble Kalman filter (TTS-EnKF) for state estimation of the nonlinear singularly perturbed system,
\item A novel two-time scale filtering methodology is developed based on TTS-EnKF for predicting the condition of the system health indicators,
\item A quantitative measure on computational complexity of our proposed TTS-EnKF estimation/prediction strategies are obtained based on the equivalent flop (EF) metric,
\item Our proposed TTS-EnKF methodology is applied and implemented for health monitoring and prognosis of a highly nonlinear industrial system, namely a gas turbine engine, when the system is affected by  degradation damage due to erosion in the turbine subsystem.
\end{enumerate}

Finally,  extensive simulations and case studies are conducted to compare and evaluate the performance and accuracy of the estimation and prediction results  achieved by using both the exact EnKF and the TTS-EnKF with the well-known particle filtering (PF) method to a gas turbine engine that is affected by degradation damage due to erosion.

The remainder of this paper is organized as follows. In Section \ref{II}, the problem statement is introduced. The necessary background related to our work is presented in Section \ref{III}. Our main proposed two-time scale EnKF is developed in Section \ref{IV}, where both fast and slow estimation filters are introduced. The prediction strategy that is based on our developed two-time scale EnKF is then introduced and developed  in Section \ref{IV}. In Section \ref{V}, extensive  case studies simulations are provided to demonstrate the merits of our proposed methodology in  erosion  degradation tracking and  prediction in a gas turbine engine. Finally, the paper is concluded in Section \ref{VI}.
\section{Problem Statement} \label{II}
Consider a time-invariant nonlinear singularly perturbed (NSP) system $\Sigma_{\epsilon}$ governed by,
\begin{align}\label{states_cont}
\Sigma_{\epsilon}: \left\{ \begin{array}{rcl}
\dot x_{1}(t)&=&f_1(x_1(t),x_2(t),\epsilon)\\
&+&g_1(x_1(t),x_2(t),\epsilon)\omega _1(t),\\
\epsilon \dot x_{2}(t)&=&f_2(x_1(t),x_2(t),\epsilon)\\
&+&{\epsilon} g_2(x_1(t),x_2(t),\epsilon)\omega _2(t),\\
y(t)&=&h(x_1(t),x_2(t),\epsilon)+\nu(t),
\end{array}\right.
\end{align}
where $x_1(t)\in \mathbb{R}^{n_s}$ and $x_2(t) \in \mathbb{R}^{n_f}$ denote the slow and fast state vectors, respectively with $x_1(t_0)=x_1(0)$ and $x_2(t_0)=x_2(0)$. The output $y(t)\in\mathbb{R}^{n_y}$ denotes the  system measurements and the parameter $0<\epsilon \ll 1$ is a sufficiently small parameter that determines the two-time scale separation of the system as $\epsilon \rightarrow 0^+$. For some $\epsilon^{\star}>0$, the functions $f_1(.),g_1(.):\mathbb{R}^{n_s}\times \mathbb{R}^{n_f}\times [0,\epsilon^{\star}) \rightarrow\mathbb{R}^{n_s},\;f_2(.),g_2(.):\mathbb{R}^{n_s}\times \mathbb{R}^{n_f}\times [0,\epsilon^{\star}) \rightarrow\mathbb{R}^{n_f}$, and $h(.):\mathbb{R}^{n_s}\times \mathbb{R}^{n_f}\times [0,\epsilon^{\star}) \rightarrow\mathbb{R}^{n_y}$ are nonlinear continuous functions. The initial conditions $x_1(0)$, and $x_2(0)$ are assumed to be deterministic \cite{socha2000exponential} and  the noise inputs $\omega_1(t)$, $\omega_2(t)$, and $\nu(t)$ are zero-mean uncorrelated noise processes with variances $Q_1(t)$, $Q_2(t)$, and $R(t)$, respectively.

The dynamical system $\Sigma_{\epsilon}$ is utilized to characterize the two-time scale property of the physical systems. One of the recent interesting applications of such modeling strategy is in damage modeling of mechanical systems as suggested in \cite{model1}. The main reason for using the singular perturbation strategy for representing the damage mechanism in physical systems is motivated by the slow dynamics (i.e., the slow changing) of the damage mechanism ($x_1(t)$ in \eqref{states_cont}) as compared to the other main physical component dynamics that are changing relatively faster ($x_2(t)$ in \eqref{states_cont}). Therefore, we utilize the model $\Sigma_{\epsilon}$ to represent the effects of  degradation damage on the health parameters of the system.

The model $\Sigma_{\epsilon}$ can be utilized to develop a unified framework for health monitoring of nonlinear systems that are assumed to be affected by degradation damage. Towards this end, the slowly time-varying health parameters of the system (which are affected by degradation) as system slow states are augmented with the system fast states. More detail regarding this formulation are provided in Subsection \ref{damage_model_definition}.

In the following section, the necessary background regarding the stochastic singular perturbation theory and the sufficient conditions that are required for its exponential stability are presented according to \cite{socha2000exponential}.
\vspace{-2mm}
\section{Background Information}\label{III}
Consider the system model $\Sigma_{\epsilon}$ where the following assumptions are held according to \cite{socha2000exponential}:\\
{\bf{Assumption 1.}} For each $\epsilon \ge 0$, $f_1(0,0,\epsilon)=0,\;f_2(0,0,\epsilon)=0,\;g_1(0,0,\epsilon)=0$, and $g_2(0,0,\epsilon)=0$.\\
{\bf{Assumption 2.}} For each $x_1(t)\in \mathbb{R}^{n_s}$, $t\ge 0$, the equation $0=f_2(x_1(t),x_2(t),0)$ has a unique solution for $x_2(t)$ denoted by $x_2^{\star}(t)=\psi(x_1(t),0)$, where $\psi(.)$ is continuously twice differentiable function.

The second assumption leads to the so-called reduced-order model (slow dynamics) that corresponds to $\Sigma_\epsilon$ by setting $\epsilon=0$ and $x_2(t)=\psi(x_1(t),0)$ in \eqref{states_cont} as follows
\vspace{-1mm}
\begin{eqnarray}\label{reduced_model}
\dot x_1(t)&=&f_1(x_1(t),\psi(x_1(t),0),0) \nonumber \\
& +& g_1(x_1(t),\psi(x_1(t),0),0)\omega_1(t).
\end{eqnarray}
Let us now define a new time variable $\tau=\frac{(t-t_0)}{\epsilon}$, as the fast-time scale or the stretched time \cite{khalil1996nonlinear} for any $t_0>0$, so that the  new states $x_{1_f}(\tau) \triangleq x_1(t_0+\epsilon \tau)=x_1(t)$ and $x_{2_f}(\tau) \triangleq x_2(t_0+\epsilon \tau)=x_2(t)$, and the noise processes $w_1(\tau)={\epsilon}\omega_1(t_0+\epsilon \tau)$ and $w_2(\tau)={\epsilon}\omega_2(t_0+\epsilon \tau)$ are obtained. Therefore, the state space representation of $\Sigma _{\epsilon}$ in terms of the new variables takes the form
\begin{align}\label{fast_eq}
\begin{split}
&\frac{{\rm{d}} x_{1_f}(\tau)}{{\rm{d}}\tau}=\epsilon f_1(x_{1_f}(\tau),x_{2_f}(\tau),\epsilon)\\
&+{\epsilon}g_1(x_{1_f}(\tau),x_{2_f}(\tau),\epsilon)w _1(\tau),\\
&\frac{{\rm{d}} x_{2_f}(\tau)}{{\rm{d}}\tau}=f_2(x_{1_f}(\tau),x_{2_f}(\tau),\epsilon) \\
&+g_2(x_{1_f}(\tau),x_{2_f}(\tau),\epsilon)w _2(\tau).
\end{split}
\end{align}
By setting $\epsilon=0$ in  \eqref{fast_eq}, we get  $\frac{{\rm{d}} x_{1_f}(\tau)}{{\rm{d}}\tau}=0$, which results in $x_{1_f}(\tau)=constant=x_{1_f}(0)=x_1(t_0)$. Therefore, the so-called boundary-layer system is given by
\begin{eqnarray}\label{boundary_eq}
\frac{{\rm{d}} x_{2_f}(\tau)}{{\rm{d}}\tau}&=&f_2(x_{1}(t_0),x_{2_f}(\tau),0)\\
&+& g_2(x_{1}(t_0),x_{2_f}(\tau),0)w _2(\tau),
\end{eqnarray}
where $x_{1}(t_0)$ is considered as a constant parameter.

We now introduce the boundary-layer or the so-called fast subsystem  $\eta(t)=x_2(t)-\psi(x_1(t),0)$. In the new fast coordinate system the singularly perturbed system $\Sigma_{\epsilon}$ can be expressed as by introducing a new fast variable as
\begin{align}\label{new_coordinate}
\begin{split}
\dot x_{1}(t)&= F_1(x_{1}(t),\eta(t),\epsilon)+ G_{11}(x_{1}(t),\eta(t),\epsilon)\omega _1(t),\\
\epsilon \dot \eta(t)&= F_2(x_{1}(t),\eta(t),\epsilon)+ G_{21}(x_{1}(t),\eta(t),\epsilon)\omega _1(t)\\
&+G_{22}(x_{1}(t),\eta(t),\epsilon)\omega _2(t),
\end{split}
\end{align}
where $x_1(t_0)=x_1(0)$ and $\eta(t_0)=x_2(t_0)-\psi(x_1(t_0),0)$, the $i$-th and the $l$-th components of $F_1,\;F_2,\;G_{11},\;G_{21}$, and $G_{22}$, for $i,j,k=1,...,n_s,\;l=1,...,n_f$ are given as follows \cite{socha2000exponential}:
$F_{1_i}(x_{1}(t),\eta(t),\epsilon)=f_{1_i}(x_{1}(t),\eta(t)+\psi(x_1(t),0),\epsilon),
F_{2_l}(x_{1}(t),\eta(t),\epsilon)=f_{2_l}(x_{1}(t),\eta(t)+\psi(x_1(t),0),\epsilon) -\epsilon[\sum^{n_s}_{j=1}\frac{\partial \psi_l(x_1(t),0)}{\partial x_{1_j}}f_{1_j}(x_1(t),\eta(t)+\psi(x_1(t),0),\epsilon)]
+\frac{1}{2}\sum_{j=1}^{n_s}\sum_{k=1}^{n_s}\frac{\partial^2\psi_l(x_1(t),0)}{\partial x_j \partial x_k}g_{1_j}(x_1(t),\eta(t)+\psi(x_1(t),0),\epsilon)g_{1_k}(x_1(t),\eta(t)+\psi(x_1(t),0),\epsilon),
G_{{11}_i}(x_1(t),\eta(t),\epsilon)=g_{1_i}(x_1(t),\eta(t)+\psi(x_1(t),0),\epsilon),
G_{{21}_l}(x_1(t),\eta(t),\epsilon)=\epsilon \sum_{j=1}^{n_s}\frac{\partial \psi_l(x_1(t),0)}{\partial x_j}g_{1_j}(x_1(t),\eta(t)+\psi(x_1(t),0),\epsilon),
G_{{22}_l}(x_1(t),\eta(t),\epsilon)={\epsilon}g_{2_l}(x_1(t),\eta(t)+\psi(x_1(t),0),\epsilon)$.

It should be noted that the reduced order slow subsystem
\begin{align}\label{reduced_y_slow}
\dot x_{1}(t)=F_1(x_1(t),0,0)+g_1(x_1(t),0,0)\omega_1(t),
\end{align}
at $\epsilon=0$ has an equilibrium at $x_1(t)=0$ and $\omega_1(t)=0$, and the boundary-layer fast subsystem is given by
\begin{eqnarray}\label{reduced_y_fast}
\frac{{\rm{d}}\eta}{{\rm{d}}\tau}=F_2(x_1(0),\eta(\tau),0)+g_2(x_1(0),\eta (\tau),0)w_2(\tau),
\end{eqnarray}
that has an equilibrium at $\eta(\tau)=0$, where $x_1(0)$ is considered as a fixed parameter.\\
\\
{\bf{Definition 1.\cite{socha2000exponential}}} Consider the nonlinear stochastic system
\begin{eqnarray}\label{def1}
\dot{x}(t)=f(t,x)+\sum^M_{i=1}g_i(t,x)\omega_i(t),\;x(t_0)=x_0
\end{eqnarray}
where $t\in\mathbb{R}^+$, $x=[x_1,...,x_n]^{\rm{T}}$ is the state vector, $f(.),\;g_i(.):\mathbb{R}^+\times \mathbb{R}^n\rightarrow\mathbb{R}^n,\;i=1,...,M$ are nonlinear deterministic vector functions as $f(.)=[f_1,...,f_n]^{\rm{T}},\;g_i=[g_{i_1},...,g_{i_n}]^{\rm{T}}$, and $\omega_i(t)$ is Gaussian noise process. Let us define the operator $\mathcal{L}^\star_{\eqref{def1}}(.)$, where the index $\eqref{def1}$ refers to the corresponding equation that this operator is applied to, as follows
\begin{eqnarray*}
\mathcal{L}^\star_{\eqref{def1}}(.)&=&\frac{\partial(.)}{\partial t}+\sum_{i=1}^n f_i(t,x)\frac{\partial(.)}{\partial {x_i}} \\ \nonumber
&+&\frac{1}{2}\sum_{i=1}^n \sum_{j=1}^n \sum_{k=1}^M g_{i_k}(t,x)g_{j_k}(t,x)\frac{\partial ^2 (.)}{\partial x_i \partial x_j}
\end{eqnarray*}

We now state the following assumptions that are necessary for introducing the Theorem 1 on the exponential stability of the system \eqref{new_coordinate} according to \cite{socha2000exponential}.\\
\\
{\bf{Assumption 3.}} A positive-definite function $V:\mathbb{R}^{n_s}\rightarrow \mathbb{R}^+$ exists which is twice continuously differentiable with respect to $x_1(t)$, and positive constants $\alpha^{\star}_{x_1}$ and $\gamma_k,\;k=1,...,4$ exist such that the following inequalities are satisfied:
\begin{eqnarray}\label{assuption3_ineq}
\begin{split}
&\gamma_1 \|x_1(t)\|^2\le V(x_1(t))\le \gamma_2 \|x_1(t)\|^2,\\
&\mathcal{L}_{\eqref{reduced_model}}^{\star}V(x_1(t))\le -2\alpha^{\star}_{x_1}V(x_1(t)),\\
&|\frac{\partial V}{\partial x_{1_i}}|\le \gamma _3 \| x \|,\; |\frac{\partial ^2 V}{\partial x_{1_i}\partial x_{1_j}}|\le \gamma_4,\; i,j=1,...,n_s.
\end{split}
\end{eqnarray}
\\
{\bf{Assumption 4.}} A positive-definite function $W:\mathbb{R}^{n_s}\times \mathbb{R}^{n_f}\rightarrow \mathbb{R}^+$ exists which is twice continuously  differentiable with respect to $\eta(t)$ and $x_1(0)$, and positive constants $\alpha^{\star}_{\eta}$ and $\nu_p,\;p=1,...,5$ exist such that the following inequalities are satisfied for $i,j=1,...,n_s$, and $k,l=1,...,n_f$:
\begin{eqnarray}\label{assuption4_ineq}
\begin{split}
&\nu_1 \|\eta(t)\|^2\le W(x_1(0),\eta(t))\le \nu _2\|\eta(t)\|^2,\\
&\mathcal{L}_{\eqref{boundary_eq}}^{\star}W(x_1(0),\eta(t))\le -2\alpha_{\eta}^{\star}W((x_1(0),\eta (t))),\\
&|\frac{\partial W}{\partial x_{1_i}}|\le \nu _3 \| \eta \|,\;|\frac{\partial W}{\partial \eta_{l}}|\le \nu _4 \|\eta  \|,\\
&|\frac{\partial ^2 W}{\partial x_{1_i}(0)\partial \eta_{k}}|\le \gamma_5, \; |\frac{\partial ^2 W}{\partial \eta_{k}\partial \eta_{l}}|\le \nu_5.
\end{split}
\end{eqnarray}
\\
{\bf{Assumption 5.}} The functions $f_1(.),\;f_2(.),\;g_1(.)$, and $g_2(.)$ are continuously differentiable with respect to $x_1$ and $x_2$, the function $\psi(x_1(t),\epsilon)$ is twice continuously differentiable with respect to $x_1$, and a real number $M_1>0$ exists such that for all $x_1 \in \mathbb{R}^{n_s}$ and $x_2 \in \mathbb{R}^{n_f}$, $i,j=1,...,n_s$, and $k,l=1,...,n_f$, we have
$|\frac{\partial f_{1_i}}{\partial x_{1_j}}|\le M_1,\;|\frac{\partial f_{1_i}}{\partial x_{2_k}}|\le M_1,\;|\frac{\partial f_{2_k}}{\partial x_{1_j}}|\le M_1,\;|\frac{\partial f_{2_k}}{\partial x_{2_l}}|\le M_1,
|\frac{\partial \psi _k}{\partial x_{1_j}}|\le M_1,\; |\frac{\partial g_{1_i}}{\partial x_{1_j}}|\le M_1,\;|\frac{\partial g_{1_i}}{\partial x_{2_l}}|\le M_1,\;|\frac{\partial g_{2_i}}{\partial x_{1_j}}|\le M_1,\;$ and $|\frac{\partial g_{2_i}}{\partial x_{2_l}}|\le M_1$.

{\bf{Assumption 6.}} The continuous functions $k_{f_1},\;k_{f_2},\;k_{g_1},\;k_{g_2}:[0,\epsilon^{\star})\rightarrow\mathbb{R}^+$, with $k_{f_1}(0)=k_{f_2}(0)=k_{g_1}(0)=k_{g_2}(0)=0$ and positive constants $d_{f_2},\;d_{g_1}$, and $d_{g_2}$ exist such that for all $x_1 \in \mathbb{R}^{n_s},\;x_2\in \mathbb{R}^{n_f}$ and $\epsilon \in (0,\epsilon^{\star}),\;i=1,...,n_s,\;l=1,...,n_f$, we have
\begin{eqnarray*}
\begin{split}
&|f_{1_i}(x_1,x_2,\epsilon)-f_{1_i}(x_1,x_2,0)|\le k_{f_1}(\epsilon)(|x_1|+|\eta|),\\
&|f_{2_l}(x_1,x_2,\epsilon)-f_{2_l}(x_1,x_2,0)|\le k_{f_2}(\epsilon)(|x_1|+|\eta|),\\
&|g_{1_i}(x_1,x_2,\epsilon)-g_{1_i}(x_1,x_2,0)|\le k_{g_1}(\epsilon)(|x_1|+|\eta|),\\
&|g_{2_l}(x_1,x_2,\epsilon)-g_{2_l}(x_1,x_2,0)|\le k_{g_2}(\epsilon)(|x_1|+|\eta|),
\end{split}
\end{eqnarray*}
where $\eta(t)=x_2(t)-\psi(x_1(t),\epsilon),\; k_{f_2}/\epsilon \le d_{f_2},\;k_{g_1}/\epsilon \le d_{g_1}$, and $k_{g_2}/\epsilon \le d_{g_2}$.\\
\\
{\bf{The Main Criterion.}} Let Assumptions 1-6 hold. Then, the positive constants $\epsilon^+,\; c$ and continuous functions $\alpha_s, \; \alpha_f:(0,\;\epsilon^{\star})$, $\phi:(0,\;\epsilon^{\star})\rightarrow \mathbb{R}^+$ exist such that the following conditions hold for $t_0>0, x_{1}(0) \in \mathbb{R}^{n_s}$ and $\eta(0) \in \mathbb{R}^{n_f}$, namely

\noindent
1) For every $\epsilon \in (0,\; \epsilon^{\star})$ and $t\ge t_0$, the solutions of \eqref{new_coordinate} are bounded as:
$\mathbb{E}|x_1(t,t_0,x_1(0),\eta(0))|\le c(|x_{1}(0)|+\phi(\epsilon)|\eta(0)|){\rm{exp}}\{-\alpha _s(t-t_0)\}; \mathbb{E}|\eta(t,t_0,x_1(0),\eta(0))|\le c|\eta(0)|{\rm{exp}}\{-\frac{\alpha_f(\epsilon)}{\epsilon}(t-t_0)\}
+\epsilon(|x_{1}(0)|+\phi(\epsilon)|\eta(0)|){\rm{exp}}\{-\alpha _s(t-t_0)\}$,
and\\
2) $\lim_{\epsilon \rightarrow 0}\alpha_s(\epsilon)=\alpha_{x_1}$, $\lim_{\epsilon \rightarrow 0}\alpha_f(\epsilon)=\alpha_{\eta}$, and $ \lim_{\epsilon \rightarrow 0}\phi(\epsilon)=0$.

\begin{theorem}
\cite{socha2000exponential} If  Assumptions 1-6 hold and for any positive $\alpha_{x_1}<\alpha^{\star}_{x_1}$, a positive constant $\epsilon^+$ and a positive continuous function $\alpha_s:(0,\;\epsilon^+)\rightarrow\mathbb{R}^+$ exist such that for every $\epsilon\in (0,\epsilon^+)$, the full-order system \eqref{new_coordinate} is exponentially stable with the rate $\alpha_s(\epsilon)$ and $\lim_{\epsilon \rightarrow 0}\alpha_s(\epsilon)=\alpha_{x_1}$, and the gain of the full-order system exponential convergence  remains finite.
\end{theorem}

Determining an explicit and exact solution to $\psi(x_1(t),\epsilon)$ is quite challenging  in general. However, by using the Gr\"{o}bner formula, the solution to $\psi(x_1(t),\epsilon)$ can be locally computed as proposed in \cite{barbot1991using}. Therefore, a common method that is used is to consider the Taylor series expansion \cite{khalil1996nonlinear, khalil, spong1987integral} of $\psi(.)$ with respect to $\epsilon$ as
\begin{eqnarray}\label{asym_ser}
\psi(x_1(t),\epsilon)=\psi_0(x_1(t))+\epsilon \psi_1(x_1(t))+O(\epsilon^2).
\end{eqnarray}

By substituting $\psi(.)$ for $x_2(t)$ in $\Sigma_{\epsilon}$ in equation (1) and applying the Assumption 2 results in the zeroth-order slow model \cite{spong1987integral} as
\begin{eqnarray}\label{slow_sys}
\dot{x}_1(t)&=&f_1(x_1(t),\psi _0(x_1(t)),0) \\ \nonumber
&+&g_1(x_1(t),\psi_0(x_1(t)),0)\omega _1(t),
\end{eqnarray}
which describes the slow dynamics of the system $\Sigma_{\epsilon}$. The solution to $x_1(t)$ from  equation \eqref{slow_sys} is now denoted by $x_{1_{s}}(t)$. The discrepancy between the response of the zeroth-order slow model \eqref{slow_sys} at $\epsilon=0$ and that of the full-order model $\Sigma_{\epsilon}$ is represented by the fast dynamics. Furthermore, one can assume that for the time interval $t\in[t_0,\;T]$ over which $x_{1_s}(t)$ exists, the following approximation is satisfied,
\begin{eqnarray}\label{slow_app}
x_1(t)=x_{1_s}(t)+O(\epsilon).
\end{eqnarray}

The second term in \eqref{asym_ser} can  now be used to specify and define a first-order slow dynamics according to
\begin{eqnarray}
\begin{split}
\dot x_1(t)&=f_1(x_1(t),\psi_0(x_1(t))+\epsilon\psi_1(x_1(t)),\epsilon)\\
&+g_1(x_1(t),\psi_0(x_1(t))+\epsilon\psi_1(x_1(t)),\epsilon)\omega_1(t).
\end{split}
\end{eqnarray}
This process can  essentially be extended  to higher order corrected slow dynamics provided a more accurate slow dynamics model is required. However, in this work, without loss of generality,  only derivations  up to $O(\epsilon ^2)$  are conducted.

To describe the dynamics of $x_2(t)$, as stated earlier, it is convenient to first define a fast time-scale  $\tau=\frac{t-t_0}{\epsilon}$, \cite{khalil1996nonlinear, khalil}, where $\tau=0$ at $t=t_0$ implies that $\eta(\tau)=x_2(\tau)-\psi_0(x_1(t))$ is well-defined. It now follows that the dynamics associated with the new fast variable $\eta(\tau)$ is governed by
\begin{eqnarray}\label{fast_app}
\frac{d\eta}{d\tau}=f_2(x_1(0),\eta(\tau)+\psi_0(x_1(0)))+O(\epsilon),
\end{eqnarray}
where $\eta(0)=x_2(0)-\psi_0(x_1(0))$. The solution to $\eta(\tau)$ from the above initial condition value problem is used as a boundary layer correction to $x_2(t)$ approximation as follows,
\begin{eqnarray}\label{boundary_app}
x_2(t)=\eta(\frac{t-t_0}{\epsilon})+\psi_0(x_1(t))+O(\epsilon).
\end{eqnarray}

In order for \eqref{boundary_app} to converge to the slow approximation of $x_2(t)=\psi_0(x_1(t))+O(\epsilon)$ (as per Assumption 2), the correction term $\eta(\tau) \rightarrow O(\epsilon)$  as $\tau \rightarrow \infty$.

In what follows, the sampled-data representation of the nonlinear  system $\Sigma_{\epsilon}$ is described  according to \cite{barbot1996analysis}, which is essential for development of our  proposed two-time scale ensemble Kalman filter (TTS-EnKF) estimation method.

It should be noted that in our proposed TTS-EnKF approach $x_{1}(t)$ is approximated and the boundary layer correction of $x_2(t)$ is performed at each time step. Therefore, Theorem 1 ensures that the error in the approximation of $x_1(t)$ and $x_2(t)$ is bounded by $O(\epsilon)$.

Before presenting our proposed TTS-EnKF approach, let us introduce the ``exact" EnKF estimation method for a full-order  nonlinear singularly perturbed system (NSP). The comparative performance of our proposed TTS-EnKF with the ``exact" EnKF are provided in subsequent sections. Consequently, a brief overview of the NSP system discritization will be presented first  below, and subsequently exact EnKF method refers to the  EnKF approach for the full-order NSP system without performing the fast-slow decomposition of the states (as done in our proposed two-time scale EnKF methodology).
\subsection{Sampled-Data Nonlinear Singularly Perturbed Systems}
In this subsection, we introduce the sampled-data representation of the  nonlinear singularly perturbed systems which is essential for developing our scheme that is based on the EnKF through an exact state estimation approach. In the exact EnKF approach to address the estimation of the fast and slow states of the NSP system, the estimation is performed without using the slow and fast states decomposition. It should also be noted that EnKF method is only applicable to discrete-time systems \cite{evensen2003}, and this is the main reason that the EnKF is not developed for the continuous-time NSP systems.

Consider the continuous-time nonlinear singularly perturbed system  governed by $\Sigma_{\epsilon}$. Assume $\iota$ denotes a sampling period where  the following conditions are satisfied,
\begin{eqnarray}
\begin{split}
\omega _i(t)&:=\omega_{i_k},\; k\iota\le t < (k+1)\iota,\;i=1,2,\\
\nu(t)&:=\nu_k,\; k\iota \le t < (k+1)\iota.
\end{split}
\end{eqnarray}
The discrete-time representation of $\Sigma_{\epsilon}$ is approximated as in \cite{barbot1996analysis, barbot1991} by the following definition.\\
\\
{\bf{Definition 2.}}
Assume that the fixed sampling period $\iota$ is sufficiently close to $\epsilon$, such that one can express $\iota$ as $\iota=\alpha \epsilon$, where $\alpha$ is a real number close to one. The fast sampled-data model  is given by
\begin{align}\label{D_z}
\begin{split}
D_{\epsilon}: \left\{ \begin{array}{rcl}
 x_{1_k}&=&x_{1_{k-1}}+\epsilon(\alpha f_1(x_{1_{k-1}},x_{2_{k-1}},\epsilon)+\alpha g_1\\
 & & (x_{1_{k-1}},x_{2_{k-1}},\epsilon)\omega _{1_k}+O(\alpha ^2))+O(\epsilon ^2),\\
 x_{{2}_{k}}&=&x_{2_{k-1}}+\alpha f_2(x_{1_{k-1}},x_{2_{k-1}},\epsilon) +\alpha \epsilon g_2\\
 & &(x_{1_{k-1}},x_{2_{k-1}},\epsilon)\omega _{1_k}+O(\alpha ^2)+O(\epsilon),\\
 y_k&=&h(x_{1_k},x_{2_k},\epsilon)+\nu _k,
\end{array}\right.
\end{split}
\end{align}
where the error due to the higher-order approximation of the system dynamics is also incorporated in the $O(\epsilon^2)$ term in $x_{1_{k}}$ and $O(\epsilon)$ term in $x_{2_{k}}$.

The discrete-time dynamical model $D_{\epsilon}$ is utilized in the remainder of this work for design of the state estimation and prediction schemes that are based on the exact EnKF approach. Towards this end, a general overview on the theory of EnKF is provided in the next subsection.
\subsection{An Overview on Ensemble Kalman Filter (EnKF) Theory}

As stated earlier, the EnKF is a suboptimal nonlinear estimation methodology where by utilizing the Monte Carlo integration, the Fokker Planck equation is approximately solved \cite{gillijns2006}. Consider a general discrete-time nonlinear system governed by the following dynamics
\begin{eqnarray}
\begin{split}
&x_{k+1}=f(x_k)+\omega _k,\\
&y_k=h(x_k)+\nu _k,
\end{split}
\end{eqnarray}
where $x_k,\; \omega_k\in \mathbb{R}^n,\;y_k,$ and $\nu_k\in \mathbb{R}^p$. The zero-mean white noise processes $\omega_k$ and $\nu_k$ have covariance matrices $Q_k$ and $R_k$, respectively. The ensemble Kalman filter (EnKF) methodology is based on two main steps that are designated as the {\it{a priori}} state estimation ({\it{forecast}}) step and the {\it{a posteriori}} state estimation ({\it{analysis}}) step \cite{gillijns2006}.

\underline{First}, at the  instant $k$, we generate $N$ ensemble members from the forecasted ({\it{a priori}}) state estimates with a random sample error that is generated from a normal distribution with the covariance $Q_k$, where the ensembles are denoted by $\{\hat {x}^{(i)}_{k|k-1},\; i=1,...,N\}$ and  generated from,
\begin{eqnarray}\label{forecast}
\hat{x}_{k|k-1}^{(i)}=f(\hat x_{k-1|k-1}^{(i)})+\omega _k^{(i)},
\end{eqnarray}
where $i=1,...,N$ refers to the ensemble number, $\hat{x}_{k|k-1}^{(i)}$ denotes the $i$-th ensemble member in the forecast step, $\hat x_{k-1|k-1}^{(i)}$ denotes the estimated ensemble member from the previous analysis step, and $\omega^{(i)}_k$ denotes the  samples from a normal distribution with the covariance $Q_k$. Note that the sample error covariance matrix which is calculated from the members of $\omega ^{(i)}_k$ converges to $Q_k$ as $N\rightarrow \infty$.

The ensemble mean $\hat{\bar{x}}_{k|k-1}$ is defined as the most probable  estimate of the state according to the Gaussian probability distribution function (as in  classical Kalman filters), as
\begin{eqnarray}
{\hat{\bar{x}}}_{k|k-1}=\frac{1}{N}\sum_{i=1}^N{\hat{x}}_{k|k-1}^{(i)}.
\end{eqnarray}

The main idea in the EnKF is to replace the error covariance matrix in the state estimation process with the ensemble covariance matrix, since the actual value of the state $x_k$ is not actually known. Therefore, the so-called {\it{a priori}} ensemble perturbation matrix $E_{k|k-1}\in \mathbb{R}^{n\times N}$ around the ensemble mean is defined as
\begin{eqnarray}\label{E-def}
E_{x_{k|k-1}}=[{\hat{x}}^{(1)}_{k|k-1}-{\hat{\bar{x}}}_{k|k-1},...,{\hat{x}}^{(N)}_{k|k-1}-{\hat{\bar{x}}}_{k|k-1}]^{\rm{T}},
\end{eqnarray}
and the output ensembles, their mean, and their ensemble perturbation matrix are accordingly computed as
\begin{eqnarray}
\begin{split}
&{\hat{y}}^{(i)}_{k|k-1}=h({\hat{x}}_{k|k-1}^{(i)}),\\
&{\hat{\bar{y}}}_{k|k-1}=\frac{1}{N}\sum_{i=1}^N{\hat{y}}_{k|k-1}^{(i)},\\
&E_{y_{k|k-1}}=[{\hat{y}}^{(1)}_{k|k-1}-{\hat{\bar{y}}}_{k|k-1},...,{\hat{y}}^{(N)}_{k|k-1}-{\hat{\bar{y}}}_{k|k-1}]^{\rm{T}}.
\end{split}
\end{eqnarray}
Next,  the covariance matrices $P^{xx}_{k|k-1}$, $P^{yy}_{k|k-1}$, and $P^{xy}_{k|k-1}$ are approximated by $\hat P^{xx}_{k|k-1}$, $\hat P^{yy}_{k|k-1}$, and $\hat P^{xy}_{k|k-1}$, as
\begin{eqnarray}
\begin{split}
&\hat P^{xx}_{k|k-1}\triangleq \frac{1}{N-1}E_{x_{k|k-1}}E_{x_{k|k-1}}^{\rm{T}},\\
&\hat P^{xy}_{k|k-1}\triangleq \frac{1}{N-1}E_{x_{k|k-1}}E_{y_{k|k-1}}^{\rm{T}},\\
&\hat P^{yy}_{k|k-1}\triangleq \frac{1}{N-1}E_{y_{k|k-1}}E_{y_{k|k-1}}^{\rm{T}}.
\end{split}
\end{eqnarray}

In fact, the ensemble members mean is interpreted as the best forecast estimate of the state, and the spread of the ensemble members around the ensemble mean is assumed to be the error between the best estimate and the actual value of the state (which is assumed to be unknown) \cite{gillijns2006, evensen2009}.

In the \underline{second} step of the EnKF algorithm, which is known as the {\it{ analysis step}} or {\it{a posteriori}} state estimation step in classical Kalman filter, the error between the observed measured outputs and the estimated outputs from the {\it{forecast step}} is utilized to reduce the error covariance of the {\it{a posteriori}} estimated state by applying the Kalman gain according to,
\begin{eqnarray}
\begin{split}
{\hat{x}}^{(i)}_{k|k}={\hat{x}}^{(i)}_{k|k-1}+{\hat{K}}_{k}(y_k-{\hat{y}}^{(i)}_{k|k-1}),
\end{split}
\end{eqnarray}
where the Kalman gain ${\hat{K}}_{k}$ is defined as
\begin{eqnarray}
\begin{split}
{\hat{K}}_{k}=\hat P^{xy}_{k|k-1}(\hat P^{yy}_{k|k-1}+R_k)^{-1}.
\end{split}
\end{eqnarray}
\underline{Finally}, the {\it{a posteriori}} error covariance matrix is approximated according to,
\begin{eqnarray}
\begin{split}
\hat P^{xx}_{k|k}\triangleq \frac{1}{N-1}E_{x_{k|k}}E_{x_{k|k}}^{\rm{T}},
\end{split}
\end{eqnarray}
where $E_{x_{k|k}}$ is defined in \eqref{E-def} with ${\hat{x}}^{(i)}_{k|k-1}$ replaced by ${\hat{x}}^{(i)}_{k|k}$ and $\hat{{\bar{x}}}_{k|k-1}$ replaced by the mean of the analysis estimate ensemble members, $\hat{{\bar{x}}}_{k|k}$.

It should be pointed out that the perturbed observation concept can also be used in the analysis step  to generate {\it{a posteriori}} ensembles \cite{evensen2003}. This method takes advantage of parallel data assimilation cycles, where for $i=1,...,N$, the {\it{a posteriori}} ensembles are updated according to
\begin{eqnarray}
\begin{split}
{\hat{x}}_{k|k}={\hat{x}}^{(i)}_{k|k-1}+{\hat{K}}_{k}(y^{(i)}_k-{\hat{y}}^{(i)}_{k|k-1}),
\end{split}
\end{eqnarray}
where $y^{(i)}_k$ denotes the perturbed observations given by
\begin{eqnarray}\label{perturbed_y}
\begin{split}
y^{(i)}_k=y_k+\nu^{(i)}_k,
\end{split}
\end{eqnarray}
where $\nu_k^{(i)}$ is a zero-mean random variable with a normal distribution and covariance $R_k$. The sample error covariance matrix that is computed from $\nu^{(i)}_k$ converges to $R_k$ as $N\rightarrow \infty$.

We are now in a position to propose and develop our proposed two-time scale estimation algorithm that is based on the EnKF for the nonlinear singularly perturbed system $D_{\varepsilon}$ in the next section.
\section{The Two-Time Scale Ensemble Kalman Filters (TTS-EnKF) for State Estimation} \label{IV}
Our proposed TTS-EnKF strategy for state estimation of the NSP system $\Sigma_{\epsilon}$ is based on the decomposition of the fast and slow dynamics of the system according to Section \ref{III}.

\subsection{Slow State EnKF Estimation}
The reduced order slow dynamics  can be approximated by
\begin{eqnarray}\label{slow_state}
\dot{x}_1(t)&=&f_1(x_1(t),\psi _0(x_1(t)),0) \\ \nonumber
&+&g_1(x_1(t),\psi_0(x_1(t),0)\omega _1(t).
\end{eqnarray}
For designing the EnKF corresponding to the slow  system, first its dynamics given by  \eqref{slow_state} is discretized. This  leads to the following discrete-time slow model
{{
\begin{eqnarray}\label{slow_state_dis}
\begin{split}
x_{1_k}&=x_{1_{k-1}}+\iota f_1(x_{1_{k-1}},\psi_0(x_{1_{k-1}}),0)\\
&+\iota g_1(x_{1_{k-1}},\psi_0(x_{1_{k-1}}),0)\omega _{1_k},
\end{split}
\end{eqnarray}
}}
The slow states of the system (that effectively correspond to the first $n_s$ largest eigenvalues of $P_{k|k-1}$), are estimated by our proposed slow filter through two main steps, namely, the time update and the measurement update steps as follows:\\

{\underline{1) Time update for the slow filter}}: Time update step is achieved  through the following procedure:
\begin{itemize}
\item {\it{A priori}} state ensemble members are generated by,
\begin{eqnarray}\label{time_update_slow}
\hat x^{(i)}_{1_{k|k-1}}&=&\hat x^{(i)}_{1_{k-1|k-1}}+ \iota f_1(\hat{x}^{(i)}_{k-1|k-1}, \nonumber \\
& &\psi _0(\hat{x}^{(i)}_{1_{k-1|k-1}}),0)+\iota g_1(\hat x^{(i)}_{1_{k-1|k-1}},\nonumber \\
& &\psi_0(\hat{x}^{(i)}_{1_{k-1|k-1}}),0)\omega^{(i)}_{1_k}
\end{eqnarray}
where $\hat{x}^{(i)}_{1_{k-1|k-1}}$ denotes the $i$-th ensemble member of the slow state in the previous time step, $\psi_0(\hat{x}^{(i)}_{1_{k-1|k-1}})$ denotes the $i$-th ensemble member of the approximated fast state that is obtained from the reduced-order model.
\item {\it{A priori}} ensemble perturbation is generated from,
\begin{eqnarray}\label{slow_apriori}
\begin{split}
\hat{\bar x}_{1_{k|k-1}}&=\frac{1}{N}\sum^N_{i=1}\hat{x}^{(i)}_{k|k-1},\\
\delta {\hat{x}}^{(i)}_{1_{k|k-1}}&=\hat x^{(i)}_{1_{k|k-1}}-\hat{\bar x}_{1_{k|k-1}},\;i=1,...,N.
\end{split}
\end{eqnarray}
\item {\it{A priori}} error covariances are computed according to,
\begin{eqnarray*}
\begin{split}
\hat{X}_{1_{k|k-1}}&=\frac{1}{\sqrt{N-1}}[\delta {\hat{x}}^{(1)}_{1_{k|k-1}},...,\delta {\hat{x}}^{(N)}_{1_{k|k-1}}]^{\rm{T}},\\
\check{P}^{\rm{s}}_{1_{k|k-1}}&=\hat{X}_{1_{k|k-1}}\hat{X}_{1_{k|k-1}}^{\rm{T}},
\end{split}
\end{eqnarray*}
where the covariance matrix $\check{P}^{\rm{s}}_{k|k-1}$ corresponds to the first $n_s$ largest eigenvalues of the exact covariance matrix  $P_{1_{k|k-1}}$, for which the fast eigenvalues of $P_{k|k-1}$ that satisfy $\frac{\lambda_k^{(n_s+j)}}{\lambda_k^{(n_s)}}\le \epsilon$, for $j=1,...,n_f$, are ignored.
\end{itemize}
{\underline{2) Measurement update for the slow filter}}: For the measurement update step as the observations become available at the time instant $k$, the {\it{a posteriori}} state estimates of the first $n_s$ slow states are obtained. In this step, the output equation in $D_\epsilon$ is replaced by its Taylor series expansion with respect to $\epsilon$ after substituting $x_{1_k}$ and $x_{2_k}$ with $\hat{x}^{(i)}_{1_{k|k-1}}$ and $\psi_0(\hat{x}^{(i)}_{1_{k-1|k-1}})$, respectively. Therefore, the zeroth-order output ensembles are computed according to
\begin{eqnarray}\label{output_expansion}
\begin{split}
\hat{y}^{(i)}_{k|k-1} = h_0(\hat{x}^{(i)}_{1_{k|k-1}},\psi_0(\hat{x}^{(i)}_{1_{k-1|k-1}}),0).
\end{split}
\end{eqnarray}

Following the above steps, the measurement ensemble perturbation matrix is obtained from
\begin{eqnarray}\label{measurement_ensemble_slow}
\begin{split}
\hat{Y}_{k|k-1}=\frac{1}{N-1}[\delta{\hat{y}}_{k|k-1}^{(1)},...,\delta{\hat{y}}_{k|k-1}^{(i)}]^{\rm{T}},
\end{split}
\end{eqnarray}
where $\delta{\hat{y}}_{k|k-1}^{(i)}={\hat{y}}_{k|k-1}^{(i)}-\frac{1}{N}\sum _{i=1}^N{\hat{y}}_{k|k-1}^{(i)}$.

Consequently, the output prediction error ensembles $\tilde y^{(i)}_{k|k-1}$ are obtained from,
\begin{eqnarray}\label{pred_err}
\begin{split}
\tilde y^{(i)}_{k|k-1} = y_k- h_0(\hat{x}^{(i)}_{1_{k|k-1}},\psi_0(\hat{x}^{(i)}_{1_{k-1|k-1}}),0).
\end{split}
\end{eqnarray}
Furthermore, the {\it{a posteriori}} ensemble members corresponding to slow states and their most probable {\it{a posteriori}}  estimates are obtained from the slow filter dynamics as
\begin{eqnarray}\label{slow_posteriori_1}
\vspace{-5mm}
\begin{split}
\hat{x}_{1_{k|k}}^{(i)}&=\hat{x}_{1_{k|k-1}}^{(i)}+\check{K}^{\rm{s}}_k \tilde y^{(i)}_{k|k-1},\\
\hat{\bar{x}}_{1_{k|k}}&=\hat{\bar{x}}_{1_{k|k-1}}+\check{K}^{\rm{s}}_k \tilde{\bar{y}}_{k|k-1},
\end{split}
\vspace{-5mm}
\end{eqnarray}
where $\tilde{\bar{y}}_{k|k-1}=\frac{1}{N}\sum^N_{i=1}\tilde{y}^{(i)}_{k|k-1}$ and $\check{K}^{\rm{s}}_k \in \mathbb{R}^{n_s\times n_y}$ denotes the Kalman gain of the slow filter. In order to select the Kalman gain for the slow filter, the \textit{a posteriori} slow filter error covariance  is defined by the following definition.
\\
{\bf{Definition 3.}} The covariance matrices $\check{P}^{xy}_{k|k-1}$ and $\check{P}^{yy}_{k|k-1}$ associated with the EnKF are approximated by $\check{P}^{xy}_{k|k-1}=\hat{X}_{1_{k|k-1}}\hat{Y}_{k|k-1}^{\rm{T}}$ and $\check{P}^{yy}_{k|k-1}=\hat{Y}_{k|k-1}\hat{Y}_{k|k-1}^{\rm{T}}$, respectively. Furthermore,  the {\textit{a posteriori}} estimation error is defined as $\tilde{x}_{1_{k|k}}=x_{1_k}-\hat{\bar{x}}_{1_{k|k}}$, so that the {\it{a posteriori}} error covariance matrix of the slow filter can be expressed as $\check{P}^{\rm{s}}_{k|k}=\mathbb{E}[\tilde{x}_{1_{k|k}}\tilde{x}_{1_{k|k}}^{\rm{T}}]$, where associated with the EnKF it is approximated by $\check{P}^{\rm{s}}_{k|k}=\hat{X}_{1_{k|k}}\hat{X}_{1_{k|k}}^{\rm{T}}$, where $\hat{X}_{1_{k|k}}$ corresponds to the ensemble perturbation matrix that is generated from the {\it{a posteriori}} estimation of the ensemble members.

Consequently, the following lemma which is inspired from  \cite{teixeira2008gain} is utilized to obtain and select the Kalman gain corresponding to the slow filter.

\vspace{2mm}
{\bf{Lemma 1.}}{\it{ Consider the cost function defined as $J_k(\check{K}^{\rm{s}}_k)=\mathbb{E}[\tilde{x}_{1_{k|k}}^{\rm{T}} W_k\tilde{x}_{1_{k|k}}]={\rm{trace}}(\check{P}^{\rm{s}}_{k|k}W_k)$, where $W_k$ denotes a positive definite matrix, $\tilde{x}_{1_{k|k}}$ denotes the {\it{a posteriori}} estimation error, and $\check{P}^{\rm{s}}_{k|k}$ denotes the {\it{a posteriori}} error covariance matrix corresponding to the slow system dynamics. The Kalamn gain $\check{K}^{\rm{s}}_k$ that minimizes this function is obtained as $\check{K}^{\rm{s}}_k=\check{P}^{xy}_{k|k-1}(\check{P}^{yy}_{k|k-1}+R_k)^{-1}$, where $\check{P}^{xy}_{k|k-1}=\hat{X}_{1_{k|k-1}}\hat{Y}_{k|k-1}^{\rm{T}}$ and $\check {P}^{yy}_{k|k-1}=\hat{Y}_{k|k-1}\hat{Y}_{k|k-1}^{\rm{T}}$.}}

\vspace{2mm}
{\bf{Proof:}} Let us follow the same procedure as in the classical Kalman filter to design the gain. All the covariance matrices in the Kalman gain are replaced by their equivalent covariance matrices that are approximated through the EnKF approach.

Assume that in the classical Kalman filter the {\it{a posteriori}} state estimation error is obtained from
\begin{eqnarray}
\begin{split}
\tilde{x}_{1_{k|k}}&=x_{1_k}-\hat{x}_{1_{k|k-1}}+\check{K}^{\rm{s}}_k(\hat{y}_{k|k-1}-y_k),\\
&= \tilde{x}_{1_{k|k-1}}+\check{K}^{\rm{s}}_k(\hat{y}_{k|k-1}-y_k),
\end{split}
\end{eqnarray}
therefore, the covariance of the {\it{a posteriori}} state estimation error is obtained from
\begin{eqnarray}\label{kalman1}
\begin{split}
\mathbb{E}\{\tilde{x}_{1_{k|k}}\tilde{x}_{1_{k|k}}^{\rm{T}}\}&= \mathbb{E}\{(\tilde{x}_{1_{k|k-1}}+\check{K}^{\rm{s}}_k(\hat{y}_{k|k-1}-y_k))\times\\
&(\tilde{x}_{1_{k|k-1}}+\check{K}^{\rm{s}}_k(\hat{y}_{k|k-1}-y_k))^{\rm{T}}\}.
\end{split}
\end{eqnarray}
Now, by expanding \eqref{kalman1} one gets
\begin{align}\label{kalman2}
\begin{split}
&J(\check{K}^{\rm{s}}_k)=\mathbb{E}\{\tilde{x}_{1_{k|k}}\tilde{x}_{1_{k|k}}^{\rm{T}}\}= \mathbb{E}\{\tilde{x}_{1_{k|k-1}}\tilde{x}_{1_{k|k-1}}^{\rm{T}}\}\\
&+\mathbb{E}\{\check{K}^{\rm{s}}_k(\hat{y}_{k|k-1}-y_k)\tilde{x}_{1_{k|k-1}}^{\rm{T}}\}
+\mathbb{E}\{\tilde{x}_{1_{k|k-1}}(\hat{y}_{k|k-1} \\
&-y_k)^{\rm{T}}\check{K}^{{\rm{s}}^{\rm{T}}}_k\}+\mathbb{E}\{ \check{K}^{\rm{s}}_k(\hat{y}_{k|k-1}-y_k)(\hat{y}_{k|k-1}-y_k)^{\rm{T}}\check{K}^{{\rm{s}}^{\rm{T}}}_k\}\\
&=\mathbb{E}\{\tilde{x}_{1_{k|k-1}}\tilde{x}_{1_{k|k-1}}^{\rm{T}}\}+\mathbb{E}\{\check{K}^{\rm{s}}_k(\hat{y}_{k|k-1}-y_k)\tilde{x}_{1_{k|k-1}}^{\rm{T}}\}\\
&+\mathbb{E}\{\tilde{x}_{1_{k|k-1}}(\hat{y}_{k|k-1}-y_k)^{\rm{T}}\check{K}^{{\rm{s}}^{\rm{T}}}_k\}\\
&-\mathbb{E}\{ \check{K}^{\rm{s}}_k(y_k-\hat{y}_{k|k-1})(y_k-\hat{y}_{k|k-1})^{\rm{T}}\check{K}^{{\rm{s}}^{\rm{T}}}_k\})
\end{split}
\end{align}
By taking the derivative of \eqref{kalman2} in terms of the Kalman gain $\check{K}^{\rm{s}}_k$ and considering that the output process $y_k$ is independent of the estimated state and measurement process, i.e. $\mathbb{E}\{\tilde{x}_{1_{k|k-1}}y_k^{\rm{T}}\}=0$ and $\mathbb{E}\{{\hat y}_{k|k-1}y_k^{\rm{T}}\}=0$, and also by noting that the covariance of the measurement noise is defined as $\mathbb{E}\{y_ky^{\rm{T}}_k\}=R_k$, it now follows that
\begin{align}\label{kalman3}
\begin{split}
&\frac{\partial J(\check{K}^{\rm{s}}_k)}{\partial \check{K}^{\rm{s}}_k}=0\Longrightarrow \frac{\partial}{\partial \check{K}^{\rm{s}}_k}(\mathbb{E}\{\tilde{x}_{1_{k|k-1}}\tilde{x}_{1_{k|k-1}}^{\rm{T}}\}\\
&+\mathbb{E}\{\check{K}^{\rm{s}}_k\hat{y}_{k|k-1}\tilde{x}_{1_{k|k-1}}^{\rm{T}}\}
+\mathbb{E}\{\tilde{x}_{1_{k|k-1}}\hat{y}_{k|k-1}^{\rm{T}}\check{K}^{{\rm{s}}^{\rm{T}}}_k\}\\
&-\mathbb{E}\{\check{K}^{\rm{s}}_k\hat{y}_{k|k-1}\hat{y}_{k|k-1}^{\rm{T}}\check{K}^{{\rm{s}}^{\rm{T}}}_k\}-\check{K}^{\rm{s}}_k R_k\check{K}^{{\rm{s}}^{\rm{T}}}_k)=0\\
&\Longrightarrow \check{K}^{\rm{s}}_k=\mathbb{E}\{\tilde{x}_{1_{k|k-1}}\hat{y}_{k|k-1}^{\rm{T}}\}(\mathbb{E}\{ \hat{y}_{k|k-1}\hat{y}_{k|k-1}^{\rm{T}}\}+R_k)^{-1},
\end{split}
\end{align}
where $\mathbb{E}\{\tilde{x}_{1_{k|k-1}}\hat{y}_{k|k-1}^{\rm{T}}\}=\check{P}^{xy}_{k|k-1}$ and $\mathbb{E}\{ \hat{y}_{k|k-1}\hat{y}_{k|k-1}^{\rm{T}}\}=\check{P}^{yy}_{k|k-1}$, which are obtained in the EnKF scheme as $\check{P}^{xy}_{k|k-1}=\hat{X}_{1_{k|k-1}}\hat{Y}_{k|k-1}^{\rm{T}}$ and $\check{P}^{yy}_{k|k-1}=\hat{Y}_{k|k-1}\hat{Y}_{k|k-1}^{\rm{T}}$. This completes the proof of the lemma.$\;\;\;\;\;\;\;\;\;\;\;\;\;\;\;\;\;\;\;\;\;\;\;\;\;\;\;\;\;\;\;\;\;\;\;\;\;\;\;\;\;\;\;\;\;\;\;\;\;\;\;\;\;\;\;\;\;\;\;\;\;\;\;\;\;\;\;\;\;\;\blacksquare$

Finally, the main three steps that are required in the measurement update of the slow states are summarized as follows:\\
(i) Measurement ensemble members and ensemble perturbation matrices are obtained according to \eqref{measurement_ensemble_slow},\\
(ii) Kalman gain selection is accomplished from Lemma 1,\\
(iii) {\it{A posteriori}} state estimations are obtained from \eqref{slow_posteriori_1}.

\subsection{Fast State EnKF Estimation}
In the next step of the estimation algorithm, the NSP system fast states are updated by assuming that the slow states are considered to be constants at their initial values at $k-1$, i.e. $\hat{\bar x}_{k-1|k-1}$ for the time interval $[k-1,k)$.
Next, the same approach that is based on the EnKF is applied  to obtain approximation to the NSP fast system states.

To design our fast  filter, we set $\epsilon=0$ in (3) to obtain
\begin{align}\label{fast_filter_eq}
\begin{split}
x_{1_f}(\tau)&= x_{1_f}(0)=x_1(t_0),\\
\frac{dx_{2_f}}{d\tau}(\tau)&= f_2(x_{1_f}(0),x_{2_f}(\tau),0)+ g_2(x_{1_f}(0),x_{2_f}(\tau),0)w _2(\tau),
\end{split}
\end{align}
Consequently, the discrete-time model of \eqref{fast_filter_eq} is now utilized in design of the EnKF as
\begin{eqnarray*}\label{fast_filter_eq_disc}
\begin{split}
 x_{2_{f_k}}&= x_{2_{f_{k-1|k-1}}}+\iota f_2(\hat{\bar x}_{1_{k-1|k-1}},x_{2_{f_{k-1|k-1}}})\\
 &+\iota g_2(\hat{\bar x}_{1_{k-1|k-1}},x_{2_{f_{k-1|k-1}}})w_{2_k}.
\end{split}
\end{eqnarray*}
The fast  state estimation filter is developed using two main steps, namely time update and  measurement update steps.
\\

{\underline{3) Time Update of the Fast Filter}}:
 The time update is performed in an $n_f$-dimensional space as follows.
\begin{enumerate}
\item {\it{A priori}} fast states ensemble generation from,
{{
\begin{align*}
\begin{split}
x_{2_{f_k}}^{(i)}&= x_{2_{f_{k-1|k-1}}}^{(i)}+\iota f_2(\hat{\bar x}_{1_{k-1|k-1}},x_{2_{f_{k-1|k-1}}}^{(i)})\\
&+\iota g_2(\hat{\bar x}_{1_{k-1|k-1}},x_{2_{f_{k-1|k-1}}}^{(i)})w_{2_k}^{(i)},
\end{split}
\end{align*}
}}
where $x_{2_{f_{k-1|k-1}}}^{(i)}$ denotes the fast ensemble members in the previous time step for $i=1,...,N$ ensembles.
\item {\it{A priori}} fast ensemble perturbation matrix generation that is obtained from,
{{
\begin{eqnarray*}
\begin{split}
\hat{\bar{x}}_{2_{f_{k|k-1}}}&=\frac{1}{N}\sum^N_{i=1}\hat{x}_{2_{f_{k|k-1}}}^{(i)},\\
\delta {\hat{x}}^{(i)}_{2_{f_{k|k-1}}}&=\hat x^{(i)}_{2_{f_{k|k-1}}}-\hat{\bar x}_{2_{f_{k|k-1}}},\;i=1,...,N,\\
\hat{X}_{2_{k|k-1}} &= \frac{1}{\sqrt{N-1}}[x_{2_{f_{k|k-1}}}^{(1)}-\hat{\bar x}_{2_{f_{k|k-1}}},...,\\
&x_{2_{f_{k|k-1}}}^{(N)}-\hat{\bar x}_{f_{2_{k|k-1}}}]^{\rm{T}}.
\end{split}
\end{eqnarray*}
}}
\end{enumerate}
In the next step, the {\it{a posteriori}} estimate of the fast states are provided and approximated.\\

{\underline{4) Measurement Update of the Fast Filter}}: It was pointed out earlier that the main assumption  used   was availability of measurements at the fast-time scale (unlike what is commonly assumed in the analysis of two-time scale systems \cite{khalil}). Therefore, the measurement update step should be performed for both the slow and fast filters. The measurement update of the fast  filter is also obtained through three main steps, namely, (i) measurement ensemble perturbation matrix computation, (ii) Kalman gain approximation, and (iii) {\it{a posteriori}} fast state estimation.\\

{\bf{Definition 4.}} The output perturbation matrix $\hat{Y}^{f}_{k|k-1}$ is defined as
$\hat{Y}^{f}_{k|k-1} = \frac{1}{\sqrt{N-1}}[\hat{y}^{f^{(1)}}_{k|k-1}-\hat {\bar{y}}^{f}_{k|k-1},...,\hat{y}^{f^{(N)}}_{k|k-1}-\hat {\bar{y}}^{f}_{k|k-1}]^{\rm{T}}$,
where $\hat{y}^{f^{(i)}}_{k|k-1}\triangleq h_0(\hat{\bar x}_{1_{k-1|k-1}},\hat{x}_{2_{f_{k|k-1}}}^{(i)}),\;i=1,...,N$, and $\hat {\bar{y}}^{f}_{k|k-1}=\frac{1}{N}\sum_{i=1}^N\hat{y}^{f^{(i)}}_{k|k-1}$.

Therefore, the following covariance matrices can be defined,
\begin{eqnarray*}
&\breve{P}^{xy}_{k|k-1} = \hat{X}_{2_{k|k-1}}\hat{Y}^{f^{\rm{T}}}_{k|k-1},\\
&\breve{P}^{yy}_{k|k-1} = \hat{Y}^{f}_{k|k-1}\hat{Y}^{f^{\rm{T}}}_{k|k-1}.
\end{eqnarray*}

The Kalman gain for the fast filter can be approximated according to  Lemma 1 and Definition 4 as,
\begin{eqnarray}
\breve{K}^f_{k}={\breve{P}}^{xy}_{k|k-1}(\breve{P}^{yy}_{k|k-1}+R_k)^{-1}.
\end{eqnarray}
Therefore, the most probable {\it{a posteriori}} fast filter state estimate is obtained from
\begin{eqnarray}\label{fast_posteriori_1}
\begin{split}
\hat{x}_{2_{f_{k|k}}}^{(i)}&=\hat{x}_{2_{f_{k|k-1}}}^{(i)}+\check{K}^{\rm{f}}_k \tilde y^{f^{(i)}}_{k|k-1},\\
\hat{\bar{x}}_{2_{f_{k|k}}}&=\hat{\bar{x}}_{2_{f_{k|k-1}}}+\check{K}^{\rm{f}}_k \tilde{\bar{y}}^{f}_{k|k-1},
\end{split}
\end{eqnarray}
where $\tilde y^{f^{(i)}}_{k|k-1}=y_k-\hat y^{f^{(i)}}_{k|k-1}$ and $\hat{\bar x}_{2_{f_{k|k}}}$ is corrected according to the received observations by applying the Kalman gain $\breve{K}^f_{k}$.

Finally, the most probable state vector $\hat{\bar x}_{k|k}$ is updated according to
  \begin{eqnarray}
  \begin{split}
  \hat{\bar x}_{k|k}&=[(\hat{\bar x}_{1_{k|k}})^{\rm{T}},\; (\hat{\bar x}_{2_{f_{k|k}}})^{\rm{T}}]^{\rm{T}}.
  \end{split}
  \end{eqnarray}

\subsection{Error Analysis of the TTS-EnKF Algorithm}
Convergence of conventional EnKF to the classical Kalman filter, and consequently to the optimal estimation of the system states for a linear problem has been addressed in \cite{mandel2011}. However, for a nonlinear  problem the convergence of EnKF to the optimal solution is not guaranteed. Therefore, for our proposed TTS-EnKF approach, the boundedness of the estimated fast and slow states are analyzed under certain conditions. Moreover,  the error due to  decomposition of the full order dynamics into slow and fast  is analyzed. First, we require the following assumption.\\
{\bf{Assumption 8.}} Consider the system $D_{\varepsilon}$ for all $\{x_{1_k},\;x_{2_k}\}$, and $p\in[1,\infty)$ such that $\| x_{1_{k}}\| _p \le b_1(k,p,\epsilon)$, $\| x_{2_{k}}\| _p \le b_2(k,p,\epsilon)$, $\| f_1(x_{1_k},x_{2_k},\epsilon)\|_p \le d_1(k,p,\epsilon)$, $\| f_2(x_{1_k},x_{2_k},\epsilon)\| _p \le d_2(k,p)$, $\| g_1(x_{1_k},x_{2_k},\epsilon)\omega_{1_k}^{(i)}\|_p \le d_3(k,p,\epsilon)$, $\| g_2(x_{1_k},x_{2_k},\epsilon)w_{2_k}^{(i)}\| _p \le d_4(k,p,\epsilon)$, $\|\psi_0(x_{1_k})\| \le c_2(k,p,\epsilon)$, and $\| h(x_{1_k},x_{2_k},\epsilon)\| _p \le d_5(k,p,\epsilon)$ are bounded for some parameters $c_i$ and $d_j$ with $i=1,2$ and $j=1,...,5$.
For a real positive number $p>1$, the norm of a vector $x \in \mathbb{R}^{n\times 1}$ is defined as $\| x\|_p:=(\sum^n_{i=1}|x_i|^p)^{\frac{1}{p}}$, with $x_i$ for $i=1,...,n$ denoting the elements of the vector $x$.

The following theorem guarantees the boundedness of the TTS-EnKF scheme {\it{a posteriori}} estimation error by considering the boundedness of the system state and output equations according to Assumption 8.
\begin{theorem}
 Consider the discrete-time nonlinear singularly perturbed system \eqref{D_z}. Let the state estimation problem be accomplished by utilizing the TTS-EnKF strategy through the {\it{a posteriori}} ensemble members update according to equations \eqref{slow_posteriori_1} and \eqref{fast_posteriori_1} for the slow and the fast states, respectively. Provided that Assumption 8 holds, there exist parameters $c_1(k,p,\varepsilon)$ and $c_2(k,p,\varepsilon)$ for all $k$ and all $p\in[1,\infty)$ such that $\| \hat{x}^{(i)}_{1_{k|k}}\|_{p} \le c_1(k,p,\varepsilon)$ and $\| \hat{x}^{(i)}_{2_{k|k}}\|_{p} \le c_2(k,p,\varepsilon)$.
\end{theorem}
{\bf{Proof:}} We invoke induction to show the result. For $k=0$, $x_{1_0}^{(i)}$ and $x_{2_0}^{(i)}$ for $i=1,...,N$ are normal distributions and bounded. Assume that for $k-1$, $\| \hat{x}^{(i)}_{1_{k-1|k-1}}\|_{p} \le c_1(k-1,p,\varepsilon)$ and $\| \hat{x}^{(i)}_{2_{k-1|k-1}}\|_{p} \le c_2(k-1,p,\varepsilon)$ for all $i$. Therefore, for the  instant $k$ associated with the slow filter  \eqref{slow_posteriori_1} we have
$\hat{x}_{1_{k|k}}^{(i)}=\hat{x}_{1_{k|k-1}}^{(i)}+\check{K}^{\rm{s}}_k \tilde y^{(i)}_{k|k-1}$.
By considering Assumption 8, from the boundedness of $\hat{x}^{(i)}_{1_{k-1|k-1}}$ and $\psi_0(\hat{x}^{(i)}_{1_{k-1|k-1}})$ one obtains $\| f_1(\hat{x}^{(i)}_{1_{k-1|k-1}},\psi_0(\hat{x}^{(i)}_{1_{k-1|k-1}})\| _p \le d_1(k-1,p,\epsilon)$ and $\| g_1(\hat{x}^{(i)}_{1_{k-1|k-1}},\psi_0(\hat{x}^{(i)}_{1_{k-1|k-1}})\omega_{1_{k-1}}^{(i)}\| _p \le d_3(k-1,p,\epsilon)$. Consequently, for the {\it{a priori}}    slow state estimate according to \eqref{time_update_slow} we have
{\small{
\begin{align}\label{slow_apriori_boundedness}
\begin{split}
\| \hat{x}^{(i)}_{1_{k|k-1}}\| _p \le c_1(k-1,p,\epsilon)+\epsilon \alpha d_1(k-1,p,\epsilon)+\epsilon \alpha d_3(k-1,p,\epsilon).
\end{split}
\end{align}
}}
Now, by considering the boundedness of $\| \hat{x}^{(i)}_{1_{k|k-1}}\| _p$ according to \eqref{slow_apriori_boundedness} and Assumption 8, the boundedness of the output equation follows since we have
\begin{eqnarray}\label{measurement_bound_slow}
\begin{split}
\| \hat{y}^{(i)}_{k|k-1}\| & \le \| h_0(\hat{x}^{(i)}_{k|k-1},\psi_0(\hat{x}^{(i)}_{k|k-1})\|\\
 &\le d_5(k-1,p,\epsilon).
\end{split}
\end{eqnarray}
Note that for computing  the Kalman gain in the measurement update step based on Lemma 1, we have $\check{K}^{\rm{s}}_k=\check{P}^{xy}_{k|k-1}(\check{P}^{yy}_{k|k-1}+R_k)^{-1}$, where $\check{P}^{xy}_{k|k-1}=\hat {X}_{1_{k|k-1}}\hat{Y}^{\rm{T}}_{k|k-1}$ and $\check{P}^{yy}_{k|k-1}=\hat{Y}_{k|k-1}\hat{Y}^{\rm{T}}_{k|k-1}$. Now, to show the boundedness of the Kalman gain we have to show the boundedness of ${\check P}^{xy}_{k|k-1}$ and $\check{P}^{yy}_{k|k-1}$, as follows
\begin{align}\label{Pxy_bound}
\begin{split}
&\| \check{P}^{xy}_{k|k-1}\| _p = \frac{1}{N-1}\|(\hat{x}^{(i)}_{1_{k|k-1}}-\hat{\bar{x}}_{1_{k|k-1}})(\hat{y}_{k|k-1}^{(i)}-\hat{\bar{y}}_{k|k-1})^{\rm{T}}\|_p \\
&=\frac{1}{N-1}\| \hat{x}^{(i)}_{k|k-1}\hat{y}^{{(i)}^{\rm{T}}}_{k|k-1}-\hat{x}^{{(i)}}_{1_{k|k-1}}\hat{\bar{y}}_{k|k-1}^{\rm{T}}-\hat{\bar{x}}_{1_{k|k-1}}\hat{y}^{{(i)}^{\rm{T}}}_{k|k-1}\\
&+\hat{\bar{x}}_{1_{k|k-1}}\hat{\bar{y}}_{k|k-1}^{\rm{T}}\| _p \\
&\le \frac{1}{N-1}(\| \hat{x}^{(i)}_{1_{k|k-1}}\hat{y}^{{(i)}^{\rm{T}}}_{k|k-1}\|_p +\|\hat{\bar{x}}_{1_{k|k-1}}\hat{\bar{y}}_{k|k-1}^{\rm{T}}\|_p).
\end{split}
\end{align}
By invoking the Cauchy inequality \cite{inequalities}, the two terms in the above inequality can be rewritten as
{{
\begin{eqnarray}\label{Pxy_bound}
\begin{split}
\| \hat{x}^{(i)}_{1_{k|k-1}}\hat{y}^{{(i)}^{\rm{T}}}_{k|k-1}\|_p &\le \mathbb{E}(\|\hat{x}^{(i)}_{1_{k|k-1}} \| ^p \|\hat{y}^{{(i)}^{\rm{T}}}_{k|k-1}\| ^p)^{\frac{1}{p}},\\
&\le \mathbb{E}(\|\hat{x}^{(i)}_{1_{k|k-1}} \| ^{2p})^{\frac{1}{2p}} \mathbb{E}(\|\hat{y}^{{(i)}^{\rm{T}}}_{k|k-1}\| ^{2p})^{\frac{1}{2p}},\\
&\le \| \hat{x}_{1_{k|k-1}}^{(i)}\|_{2p} \| \hat{y}^{{(i)}^{\rm{T}}}_{k|k-1}\|_{2p}
\end{split}
\end{eqnarray}
}}
which yields
$\| \check{P}^{xy}_{k|k-1}\| _p \le \frac{1}{N-1}(\| \hat{x}_{1_{k|k-1}}^{(i)}\|_{2p} \| \hat{y}^{{(i)}^{\rm{T}}}_{k|k-1}\|_{2p}
+\| \hat{\bar x}_{1_{k|k-1}}\|_{2p} \| \hat{\bar y}^{\rm{T}}_{k|k-1}\|_{2p}),
\le \frac{2}{N-1}c_1(k-1,p,\epsilon)d_5(k-1,p,\epsilon)$.
Similar to the derivations used in $\| \check{P}^{xy}_{k|k-1}\| _p$ for $\| \check{P}^{yy}_{k|k-1}\| _p$ we can also obtain
$\| \check{P}^{yy}_{k|k-1}\| _p \le \frac{1}{N-1}(\| \hat{y}_{k|k-1}^{(i)} \hat{y}_{k|k-1}^{{(i)}^{\rm{T}}}\|_{2p} +\| \hat{\bar y}_{k|k-1} \hat{\bar y}_{k|k-1}^{\rm{T}})\|_{2p})\le \frac{1}{N-1}(\| \hat{y}_{k|k-1}^{(i)} \|^2_{2p} +\| \hat{\bar y}_{k|k-1} \|^2_{2p})\le \frac{2}{N-1}d^2_5(k-1,p,\epsilon)$.
Now to show the boundedness of $\check K^{\rm{s}}_k$, note that $P^{yy}_{k|k-1}$ is symmetric and positive semi-definite and $R_k$ is  symmetric positive definite matrix. Therefore, we have
\begin{eqnarray}\label{Pyy_bound2}
\begin{split}
\| (\check{P}^{yy}_{k|k-1}+R_k)^{-1}\|  \le \| R_k^{-1}\|\le {\rm{cte}}(k)
\end{split}
\end{eqnarray}
where ${\rm{cte}}(k)$ denotes a constant parameter at the time instant $k$. The inequality  \eqref{Pyy_bound2} together with the bound on $\| \check{P}^{xy}_{k|k-1} \|_p$ according to \eqref{Pxy_bound} yield
\begin{eqnarray}\label{K_bound}
\begin{split}
\| \check{K}^{\rm{s}}_k \| _p &\le  \| \check{P}^{xy}_{k|k-1} \|_p cte(k)\\
&\le \frac{2}{N-1}c_1(k-1,p,\epsilon)d_5(k-1,p,\epsilon) cte(k),
 \end{split}
\end{eqnarray}
where $N$ is a sufficiently large number ($N \rightarrow \infty$).

Finally, to show the boundedness of the {\it{a posteriori}} slow state estimate, consider equation \eqref{slow_posteriori_1} that yields,
$\| \hat{x}^{(i)}_{1_{k|k}} \| _p \le \| \hat{x}^{(i)}_{1_{k|k-1}} \| _p +\| \check{K}^{\rm{s}}_k \tilde y^{(i)}_{k|k-1}\| _p  \le c_1(k-1,p,\epsilon)+\epsilon \alpha (d_1(k-1,p,\epsilon)+d_3(k-1,p,\epsilon))+\| \check{K}^{\rm{s}}_k \|_{2p}  \|\hat{y}^{{(i)}}_{k|k-1}\|_{2p} \le c_1(k-1,p,\epsilon)+\epsilon \alpha (d_1(k-1,p,\epsilon)+d_3(k-1,p,\epsilon))+\frac{2}{(N-1)}c_1(k-1,p,\epsilon)d^2_5(k-1,p,\epsilon) cte(k)$.
Hence, by applying the Jensen's inequality \cite{inequalities} for any $\hat{x}^{(i)}_{1_{k|k}}$, we obtain
$\| \frac{1}{N}\sum^N_{i=1}\hat{x}^{(i)}_{1_{k|k}}\|_p \le \frac{1}{N}\sum^N_{i=1}\| \hat{x}^{(i)}_{1_{k|k}}\|_{p}$,
which yields
{{
\begin{align}\label{x_1 apriori bound}
\begin{split}
\| \hat{\bar{x}}_{1_{k|k}}\| _p & \le c_1(k-1,p,\epsilon)+\epsilon \alpha (d_1(k-1,p,\epsilon)\\
&+d_3(k-1,p,\epsilon))+\frac{2}{(N-1)}c_1(k-1,p,\epsilon)d^2_5(k-1,\\
& p,\epsilon) cte(k) \le c_1(k,p,\epsilon).
\end{split}
\end{align}
}}

Next, the boundedness of the {\it{a posteriori}} estimate of the fast states through the fast filter is investigated. By invoking induction, we assume that $\| x^{(i)}_{1_{k-1|k-1}}\| _p \le c_1(k-1,p,\epsilon)$ and $\| x^{(i)}_{2_{k-1|k-1}}\| _p \le c_2(k-1,p,\epsilon)$. Using the same approach as in the slow filter, an upper bound on the estimated {\it{a posteriori}} fast state is obtained as
$\| \hat{\bar{x}}_{2_{k|k}}\| _p  \le c_2(k-1,p,\epsilon)+\epsilon \alpha (d_2(k-1,p,\epsilon)+d_4(k-1,p,\epsilon))+\frac{2}{(N-1)}c_2(k-1,p,\epsilon)d^2_5(k-1,p,\epsilon)
cte(k) \le c_2(k,p,\epsilon)$.
This completes the proof of the theorem.$\;\;\;\;\;\;\;\;\;\;\;\;\;\;\;\;\;\;\;\;\;\;\;\;\blacksquare$

We now investigate the boundedness of the estimation error based on the error analysis associated with the {\it{a posteriori}}  state estimation and the one that is obtained from the  reduced order system model. If we substitute $x_{2_{k-1}}$ in $D_\epsilon$ with $\psi_0(x_{1_{k-1}})$, the reduced order slow model for estimating $x_{{\rm{s}}_{1_k}}$ is obtained, where according to Lemma 1  and equation  \eqref{slow_app}, $x_{1_{k}}=x_{{\rm{s}}_{1_k}}+O(\epsilon)$. In our developed TTS-EnKF filter, $x_{{\rm{s}}_{1_k}}$ is estimated. Therefore, the estimation error is represented by
$e_{{\rm{s}}_{1_k}}=x_{{\rm{s}}_{1_k}}-\hat{\bar{x}}_{1_{k|k}}$,
where $e_{{\rm{s}}_{1_k}}$ denotes the slow filter estimation error. Hence, an upper bound on this error can be obtained as
$\| e_{{\rm{s}}_{1_k}}\| _p \le \| x_{{\rm{s}}_{1_k}} \| _p + \|
\hat{\bar{x}}_{1_{k|k}} \|_p+O(\epsilon) \le \| x_{1_{k-1}} \| _p +\epsilon \alpha (\| f_1(x_{1_{k-1}},x_{2_{k-1}})\| _p +\| g_1(x_{1_{k-1}},x_{2_{k-1}})\omega_{1_k}\| _p)
+ \epsilon O(\alpha ^2) +O(\epsilon)+\| \hat{\bar{x}}_{1_{k|k}} \| _p \le 2 c_1(k-1,p,\epsilon) + 2 \epsilon \alpha (d_1(k-1,p,\epsilon)+d_3(k-1,p,\epsilon))+\frac{2}{(N-1)}c_1(k-1,p,\epsilon)d^2_5(k-1,p,\epsilon) cte(k)+O(\epsilon)$,
where the last inequality is obtained by applying Assumption 8 and by replacing $\| \hat{\bar{x}}_{1_{k|k}} \| _p$ with the bound from \eqref{x_1 apriori bound}. An error of magnitude $O(\epsilon)$ is added due to the resulting discretization error  and considering that $\alpha$ is very close to $1$. Now, as $N\rightarrow \infty$, if $c_1(k-1,p,\epsilon)d^2_5(k-1,p,\epsilon) cte(k) \ll N$, the term $\frac{2}{(N-1)}c_1(k-1,p,\epsilon)d^2_5(k-1,p,\epsilon) cte(k)$ will tend to zero, and one can approximate $2 \epsilon \alpha (d_1(k-1,p,\epsilon)+d_3(k-1,p,\epsilon)) +O(\epsilon)$ by an $O(\epsilon)$ term. Consequently, the upper bound on the estimation error for the reduced order slow model is given by
$\| e_{{\rm{s}}_{1_k}}\| _p \le 2 c_1(k-1,p,\epsilon) + O(\epsilon)$.
Now, the error in the estimation of $x_{1_{k}}$ can be obtained as
\begin{eqnarray} \label{error_2}
\begin{split}
e_{1_k}=x_{{\rm{s}}_{1_k}}+O(\epsilon)-\hat{\bar{x}}_{1_{k|k}}.
\end{split}
\end{eqnarray}
Similarly, the upper bound on $e_{1_k}$ can be expressed as
$\| e_{1_k}\| _p \le 2 c_1(k-1,p,\epsilon) + O(\epsilon)$.

Therefore, the estimation error based on the discrepancy between the actual  $x_{2_k}$ from $D_\epsilon$ and the estimated $\hat{\bar{x}}_{2_{k|k}}$ can be expressed as $e_{2_k}=x_{2_k}-\hat{\bar{x}}_{2_{k|k}}$,
where $e_{2_k}$ denotes the estimation error of the fast filter. Hence, an upper bound on this error can be obtained as
$\| e_{2_k}\| _p \le \| x_{2_k} \| _p + \| \hat{\bar{x}}_{2_{k|k}} \|_p \le \| x_{2_{k-1}} \| _p + \alpha (\| f_2(x_{1_{k-1}},x_{2_{k-1}})\| _p+ \| g_2(x_{1_{k-1}},x_{2_{k-1}})\omega_{2_k}\| _p) + O(\alpha ^2) +\| \hat{\bar{x}}_{2_{k|k}} \| _p \le 2 c_2(k-1,p,\epsilon) + \alpha (d_2(k-1,p,\epsilon)+d_4(k-1,p,\epsilon))+\epsilon \alpha (d_2(k-1,p,\epsilon)+d_4(k-1,p,\epsilon))+\frac{2}{(N-1)}c_2(k-1,p,\epsilon)d^2_5(k-1,p,\epsilon) cte(k)+O(\alpha ^2)\le 2 c_2(k-1,p,\epsilon) + \alpha (d_2(k-1,p,\epsilon)+d_4(k-1,p,\epsilon))+O(\alpha^2)+O(\epsilon)$.
Finally, the error of the fast filter is propagated with the order of $O(\alpha)$, whereas for the slow filter it is propagated with an order of $O(\epsilon)$.

In the next section,  our developed TTS-EnKF filters will be developed for solving  the long-term prediction of the system states/health parameters in accomplishing the health monitoring problem.
\section{Prediction Scheme Based on Two-Time Scale EnKF}\label{IV}
In this section, our previously developed two-time scale EnKF scheme is utilized for performing a long-term prediction of the nonlinear system slowly time-varying health parameters as well as its fast states. This problem is generally considered as a second module in development of an integrated framework for health monitoring of complex engineering systems.
\subsection{Prediction Framework Based on Two-Time Scale EnKF}
The main challenge in the prediction problem is that the prediction errors increase as the prediction horizon is extended. This problem is directly related to lack of availability of actual observations after the time instant $k$. In other words,  information on actual observations cannot be used for reducing the resulting error covariances in the {\it{a posteriori}} state estimation process.

In our proposed framework, we follow the main idea of the EnKF which substitutes the actual states with their sequence of ensemble members to obtain the estimation error. Moreover,  we also replace  observations  in the measurement update step with  approximated observations  in both  slow and fast filters that result from approximated observations ensembles.

Our proposed prediction scheme consists of two filters, namely the slow and the fast filters for updating the health parameters as well as the states, respectively. Consequently, our proposed prediction strategy that is  based on the two-time scale EnKF scheme can be summarized as follows.

\subsection{Prediction of the Slow States}
Our prediction strategy is implemented  through two main steps (similar to the estimation strategy), namely the time update and the measurement update step (based on approximated measurements).

\subsubsection{\underline{Time Update of the Slow Filter}}
The state of this filter is defined as $x_{{\rm{s}}_{1_k}}$. The ensemble members are generated by using  the following equations for an $l>1$ step ahead prediction
\begin{align}\label{time_update_slow_pred}
\begin{split}
&\hat{x}^{{(i)}^{-}}_{1_{k+l|k}}=\hat{x}^{{(i)}^{+}}_{1_{k+l-1|k}}+\iota f_1(\hat x^{{(i)}^{+}}_{1_{k+l-1|k}},\psi _0(\hat x^{{(i)}^{+}}_{1_{k+l-1|k}}),0)\\
&+ \iota g_1(\hat x^{{(i)}^{+}}_{1_{k+l-1|k}},\psi _0(\hat x^{{(i)}^{+}}_{1_{k+l-1|k}}),0)\omega^{(i)}_{1_{k+l-1}},
\end{split}
\end{align}
where $i=1,...,N$, the superscript ${(-)}$ refers to the predicted state in the previous time step before performing the covariance correction in the measurement update step, and the superscript ${(+)}$ refers to the approximated  state after performing the measurement update.

The most probable states and their corresponding ensemble perturbations are generated according to
\begin{eqnarray}
\begin{split}
&{\hat{\bar {x}}}_{1_{k+l|k}}^{-}=\frac{1}{N}\sum_{i=1}^N \hat{x}_{1_{k+l|k}}^{{(i)}^{-}},\\
&\delta {\hat{x}}_{1_{k+l|k}}^{{(i)}^{-}}={\hat{x}}_{1_{k+l|k}}^{{(i)}^{-}}-{\hat{\bar {x}}}^{-}_{1_{k+l|k}}.
\end{split}
\end{eqnarray}
The vector $\hat{X}_{{\rm{s}}_{1_{k+l|k}}}^-=\hat{X}_{1_{k+l|k}}^{-}$ is defined as $\frac{1}{\sqrt{N-1}}[\delta {\hat{x}}^{{(1)}^{-}}_{1_{k+l|k}},...,\delta {\hat{x}}^{{(N)}^{-}}_{1_{k+l|k}}]^{\rm{T}}$.
\subsubsection{\underline{Measurement Update of the Slow Filter}}
For  measurement update, as stated earlier due to absence of observations, one requires to apply an alternative approach to reduce the prediction error in this step. We proposed to utilize the following approximation for the $l$-step ahead prediction of the observation vector, namely
\begin{eqnarray}\label{observation_pred1}
\begin{split}
y^{\rm{s}}_{k+l}\approx h_0(\hat{\bar x}^{-}_{1_{k+l|k}},\psi _0 (\hat{\bar x}^{-}_{1_{k+l|k}}),0),
\end{split}
\end{eqnarray}
where $y^{\rm{s}}_{k+l}$ denotes the predicted observations of the slow filter. Hence, the approximated observation vector according to \eqref{observation_pred1} is utilized in the slow filter to predict the system slow states.

To summarize, the prediction scheme for the slow filter is performed according to the following steps:
\begin{enumerate}
\item The output perturbation matrix is computed from $\hat{Y}^{\rm{s}}_{k+l|k}=[\delta {\hat y}^{{\rm{s}}^{(1)}}_{k+l|k},...,\delta {\hat y}^{{\rm{s}}^{(N)}}_{k+l|k}]^{\rm{T}}$, where for $i=1,..,N$, $\delta \hat {y}^{{\rm{s}}^{(i)}}_{k+l|k}=h_0(\hat{x}^{(i)^-}_{1_{k+l|k}},\psi _0 (\hat{x}^{(i)^-}_{1_{k+l|k}}),0)-\frac{1}{N}\sum_{i=1}^N h_0(\hat{x}^{(i)^-}_{1_{k+l|k}},\psi _0 (\hat{x}^{(i)^-}_{1_{k+l|k}}),0)$,
\item The Kalman gain is computed from $\check{K}^{\rm{s}}_{k+l}=\check{P}^{\rm{xy}}_{k+l|k}(\check{P}^{\rm{yy}}_{k+l|k}+R_k)^{-1}$, where $\check{P}^{\rm{xy}}_{k+l|k}=\hat{X}^-_{1_{k+l|k}}\hat{Y}^{{\rm{s}}^{\rm{T}}}_{k+l|k}$ and $\check{P}^{\rm{yy}}_{k+l|k}=\hat{Y}^{\rm{s}}_{k+l|k}\hat{Y}^{{\rm{s}}^{\rm{T}}}_{k+l|k}$.
\item The prediction of the {\it{a posteriori}} state ensemble members is computed from
\begin{eqnarray}\label{measurement_update_slow_pred}
\hat{x}^{{(i)}^+}_{1_{k+l|k}}=\hat{x}^{{(i)}^-}_{1_{k+l|k}}+\check{K}^{\rm{s}}_{k+l}\tilde y^{\rm{s}^{(i)}}_{k+l|k},
\end{eqnarray}
 where $\tilde{y}^{{\rm{s}}^{(i)}}_{k+l|k}=y^{\rm{s}}_{k+l}-h_0(\hat{x}^{(i)^-}_{1_{k+l|k}},\psi _0 (\hat{x}^{(i)^-}_{1_{k+l|k}}),0).$
\item The most probable {\it{a posteriori}} state estimate is computed from $\hat{\bar{x}}_{1_{k+l|k}}^+=\frac{1}{N}\sum^{N}_{i=1}\hat{x}^{{(i)}^+}_{1_{k+l|k}}$.
\end{enumerate}

In the next subsection, the prediction scheme for the fast states of the system is provided in detail.
\subsection{Prediction of the Fast States}
The prediction scheme for this filter is also implemented  through two main steps, namely the time update and the measurement update steps where the ensemble perturbations update is also performed in the second step.

\subsubsection{\underline{Time Update of the Fast Filter}}
For this filter, the slow states of the system are considered as fixed and equal to their most probable predicted values obtained from the previous time step, i.e., $x_{1_{k+l}}\approx{\hat{\bar{x}}}^{(+)}_{1_{k+l-1|k}}$. Therefore, the time update is performed according to the following expressions
\begin{eqnarray}\label{time_update_fast_pred}
\begin{split}
&\hat{x}_{1_{k+l|k}}={\hat{\bar{x}}}^{(+)}_{1_{k+l-1|k}},\\
&\hat{\bar x}_{2_{k+l|k}}^{(i)^{-}}=\hat{x}_{2_{k+l-1|k}}^{(i)^{+}}+\iota f_2({\hat{\bar{x}}}^{(+)}_{1_{k+l-1|k}},x_{2_{k+l-1|k}}^{(i)^{+}})\\
&+\iota g_2({\hat{\bar{x}}}^{(+)}_{1_{k+l-1|k}},x_{2_{k+l-1|k}}^{(i)^{+}})+w ^{(i)}_{2_{k+l}},
\end{split}
\end{eqnarray}
where $\hat{x}_{2_{k+l-1|k}}^{(i)^{+}}$ denotes the predicted fast ensembles from the previous time step for $i=1,...,N$ members. We also define the vector $\hat{X}^{{\rm{f}}^-}_{k+l|k}=\hat{X}_{2_{k+l|k}}^{-}=\frac{1}{\sqrt{N-1}}[\delta {\hat{x}}^{{(1)}^{-}}_{2_{k+l|k}},...,\delta {\hat{x}}^{{(N)}^{-}}_{2_{k+l|k}}]^{\rm{T}}$.
\subsubsection{\underline{Measurement Update of the Fast Filter}}
For the measurement update, similar to the slow filter, we propose to utilize an approximation for the $l$-step ahead prediction of the observation vector as follows,
\begin{eqnarray}\label{observation_pred2}
\begin{split}
y^{\rm{f}}_{k+l}\approx h_0({\hat{\bar{x}}}_{1_{k+l-1|k}}^{+},{\hat{\bar{x}}}^{-}_{2_{k+l|k}}),
\end{split}
\end{eqnarray}
where $y^{\rm{f}}_{k+l}$ denotes predicted observations from the fast filter.

Therefore, the approximated observation vector according to \eqref{observation_pred2} is used in the fast filter to predict the system fast states according to the following steps:
\begin{enumerate}
\item The output perturbation matrix is computed from $\hat{Y}^{\rm{f}}_{k+l|k}=[\delta {\hat y}^{{\rm{f}}^{(1)}}_{k+l|k},...,\delta {\hat y}^{{\rm{f}}^{(N)}}_{k+l|k}]^{\rm{T}}$, where for $i=1,..,N$, $\delta \hat {y}^{{\rm{f}}^{(i)}}_{k+l|k}=h_0(\hat{\bar x}^{+}_{1_{k+l-1|k}},\hat{x}^{(i)^-}_{2_{k+l|k}})-\frac{1}{N}\sum_{i=1}^N h_0(\hat{\bar x}^{+}_{1_{k+l-1|k}},\hat{x}^{(i)^-}_{2_{k+l|k}})$,
\item The Kalman gain is computed from $K^{\rm{f}}_{k+l}=\breve{P}^{\rm{xy}}_{k+l|k}(\breve{P}^{\rm{yy}}_{k+l|k}+R_k)^{-1}$, where $\breve{P}^{\rm{xy}}_{k+l|k}=\hat{X}^{-}_{2_{k+l|k}}\hat{Y}^{{\rm{f}}^{\rm{T}}}_{k+l|k}$, and $\breve{P}^{\rm{yy}}_{k+l|k}=\hat{Y}^{\rm{f}}_{k+l|k}\hat{Y}^{{\rm{f}}^{\rm{T}}}_{k+l|k}$.
\item The prediction of the {\it{a posteriori}} state ensemble members of fast states is computed from $\hat{x}^{{(i)}^+}_{2_{k+l|k}}=\hat{x}^{{(i)}^-}_{2_{k+l|k}}+\breve{K}^{\rm{f}}_{k+l}\tilde y^{\rm{f}^{(i)}}_{k+l|k}$, where $\tilde{y}^{{\rm{f}}^{(i)}}_{k+l|k}=y^{\rm{f}}_{k+l}-h_0({\hat{\bar{x}}}^+_{1_{k+l-1|k}},{\hat{x}}_{2_{k+l|k}}^{(i)^-}).$
\item The most probable {\it{a posteriori}} state estimate is computed from $\hat{\bar{x}}_{2_{k+l|k}}^+=\frac{1}{N}\sum^{N}_{i=1}\hat{x}^{{(i)}^+}_{2_{k+l|k}}$.
\end{enumerate}

It follows that as the step-ahead prediction horizon is extended, errors in state prediction  also increase. The following theorem which is inspired from Theorem 2, provides bounds on the state estimation results for the TTS-EnKF as a function of the $l$-step ahead prediction horizon.\\
\begin{theorem}
Let Assumption 8 and Theorem 2 results hold. The $l$-step ahead prediction error of the system slow and fast states  given in \eqref{D_z} and that utilize the TTS-EnKF scheme remains bounded with an error of the order of $(l+2)O(\epsilon)$ for $\hat{\bar x}^+_{1_{k+l|k}}$ and of the order $(l+1)O(\alpha^2)+(l+2)O(\epsilon)$ for $\hat{\bar x}^+_{2_{k+l|k}}$.
\end{theorem}
{\bf{Proof:}} Following the prediction scheme, it is known that the predicted state from the previous time step is utilized to predict the state in the next time instant. Therefore, the error due to discretization of the system as well as the fast-slow decomposition of the system states do propagate throughout the prediction scheme to  future time steps. Therefore, by  substituting back the predicted values of $\hat{x}_{1_{k+j|k}}^{(i)^-}$ and $\hat{x}_{2_{k+j|k}}^{(i)^-}$ for $j=0,...,l-1$ into equations \eqref{time_update_slow_pred} and \eqref{measurement_update_slow_pred} for the slow filter, one obtains
$\hat{x}^{{(i)}^{+}}_{1_{k+l|k}}=\hat{x}^{{(i)}^{+}}_{1_{k+l-1|k}}+\epsilon \alpha ({f}_1(\hat x^{{(i)}^{+}}_{1_{k+l-1|k}},\psi_0(\hat x^{{(i)}^{+}}_{1_{k+l-1|k}}),0)+ {g}_1(\hat x^{{(i)}^{+}}_{1_{k+l-1|k}},\psi_0(\hat x^{{(i)}^{+}}_{1_{k+l-1|k}}),0) \omega_{1_{k+l}}^{{(i)}^{+}})+\check{K}^{\rm{s}}_{k+l}\tilde y^{\rm{s}^{(i)}}_{k+l|k}=\hat{x}^{{(i)}^{+}}_{1_{k|k}}+\epsilon \alpha \sum_{j=0}^{l-1}({f}_1(\hat x^{{(i)}^{+}}_{1_{k+j|k}},\psi_0(\hat x^{{(i)}^{+}}_{1_{k+j|k}}),0)+{g}_1(\hat x^{{(i)}^{+}}_{1_{k+j|k}},\psi_0(\hat x^{{(i)}^{+}}_{1_{k+j|k}}),0)\omega ^{(i)}_{k+j+1})+\sum_{j=1}^l\check{K}^{\rm{s}}_{k+j}\tilde y^{\rm{s}^{(i)}}_{k+j|k}$.
Now, a bound on the predicted state $\hat{x}^{{(i)}^{+}}_{1_{k+l|k}}$ and the prediction error can be obtained by considering the Assumption 8 and Theorem 2 as follows:
$\| \hat{x}^{{(i)}^{+}}_{1_{k+l|k}}\| \le c_1(k,p,\epsilon)+\alpha\epsilon \sum_{j=0}^{l-1}(d_1(k+j,p,\epsilon)+d_3(k+j,p,\epsilon))
+\sum_{j=1}^l\frac{2}{(N-1)}c_1(k+j,p,\epsilon)d_5^2(k+j,p,\epsilon)cte(k+j)$, and
$\|e_{{\rm{s}}_{1_{k+l}}}\| \le c_1(k+l-1,p,\epsilon)+c_1(k,p,\epsilon)+\epsilon \alpha (d_1(k+l-1,p,\epsilon)+d_3(k+l-1,p,\epsilon))+\alpha\epsilon \sum_{j=0}^{l-1}(d_1(k+j,p,\epsilon)+d_3(k+j,p,\epsilon))
+\sum_{j=1}^l\frac{2}{(N-1)}c_1(k+j,p,\epsilon)d_5^2(k+j,p,\epsilon)cte(k+j)+O(\epsilon)$,
by assuming $l\ll N$.

Now the upper bound on the prediction error corresponding to the slow states is given by
$\|e_{{\rm{s}}_{1_{k+l}}}\| \le c_1(k+l-1,p,\epsilon)+c_1(k,p,\epsilon)+(l+2)O(\epsilon)$.
The same procedure can be applied to $\hat{x}^{{(i)}^{+}}_{2_{k+l|k}}$ to obtain the upper bound on the $l$-step ahead prediction as follows:
$\hat{x}^{{(i)}^{+}}_{2_{k+l|k}}=\hat{x}^{{(i)}^{+}}_{2_{k+l-1|k}}+ \alpha \epsilon({f}_2(\hat{\bar x}^{+}_{1_{k+l-1|k}},\hat x^{{(i)}^{+}}_{2_{k+l-1|k}})+g_2(\hat{\bar x}^{+}_{1_{k+l-1|k}},\hat x^{{(i)}^{+}}_{2_{k+l-1|k}})\omega^{(i)}_{2_{k+l}})+\check{K}^{\rm{f}}_{k+l}\tilde y^{\rm{f}^{(i)}}_{k+l|k}=\hat{x}^{{(i)}^{+}}_{2_{k|k}}+ \alpha \epsilon \sum_{j=0}^{l-1}({f}_2(\hat{\bar x}^{+}_{1_{k+j|k}},\hat x^{{(i)}^{+}}_{2_{k+j|k}})+{g}_2(\hat{\bar x}^{+}_{1_{k+j|k}},\hat x^{{(i)}^{+}}_{2_{k+j|k}})\omega^{(i)}_{2_{k+j}})+\sum_{j=1}^l\check{K}^{\rm{f}}_{k+j}\tilde y^{\rm{f}^{(i)}}_{k+j|k},\le c_2(k,p,\epsilon)+\alpha \epsilon \sum_{j=0}^{l-1}(d_2(k+j,p,\epsilon)+d_4(k+j,p,\epsilon))+
\sum_{j=1}^l\frac{2}{N-1}c_2(k+j,p,\epsilon)d_5^2(k+j,p,\epsilon)cte(k+j)$, and
$\|e_{2_{k+l}}\| \le c_2(k+l-1,p,\epsilon)+c_2(k,p,\epsilon)+ \alpha (d_2(k+l-1,p,\epsilon)+d_4(k+l-1,p,\epsilon))+O(\epsilon)+O(\alpha^2)+\alpha\epsilon \sum_{j=0}^{l-1}(d_2(k+j,p,\epsilon)+d_4(k+j,p,\epsilon))
+\sum_{j=1}^l(\frac{2}{(N-1)}c_2(k+j,p,\epsilon)d_5^2(k+j,p,\epsilon)cte(k+j)+O(\epsilon)+O(\alpha^2))\le c_2(k+l-1,p,\epsilon)+c_2(k,p,\epsilon)+ \alpha (d_2(k+l-1,p,\epsilon)+d_4(k+l-1,p,\epsilon))+(l+2)O(\epsilon)+(l+1)(\alpha^2)$.
Therefore, the upper bounds on the $l$-step ahead predicted states as well as the order of the propagated error as a function of the prediction horizon are obtained explicitly. This completes the proof of the theorem. $\;\;\;\;\;\;\;\;\;\;\;\;\;\;\;\;\;\;\;\;\;\;\;\;\;\;\;\;\;\;\;\;\;\;\;\;\;\;\;\;\;\;\;\;\;\;\;\;\;\;\;\;\;\;\;\blacksquare$

The results from  Theorem 3 show that although the $l$-step prediction of the system states remains bounded for a bounded $l$, however as the prediction horizon is extended the errors due to approximation of the exact system into slow and fast subsystems cause additional errors in the resulting predictions. Therefore, the prediction horizon should be chosen carefully such that ignoring the slow-fast decomposition errors in the developed TTS-EnKF scheme cannot be significant. In Section \ref{V}, our developed TTS-EnKF estimation/prediction strategies are applied to track the degradation phenomenon and its propagation prediction for a long-term horizon interval in a gas turbine engine system.

In addition to  prediction accuracy of our nonlinear filtering strategies  developed for NSP systems, the issue of computational cost associated with implementation of the  developed schemes is also an important issue. This problem is now investigated by determining a trade-off between the accuracy and the cost of any proposed scheme. In the next subsection, the computational cost of our developed TTS-EnKF method is quantified and is compared with the well-known particle filtering prediction  approach.
\subsection{Complexity Analysis of the TTS-EnKF Prediction Scheme} \label{Complexity}

In this subsection, the computational complexity of our proposed TTS-EnKF prediction scheme is quantitatively obtained and analyzed. The analysis is based on the number of floating-point operations (flops) that are required by each algorithm. This is commonly known as the equivalent flop (EF) analysis.
\begin{table*}
\caption{The Total Equivalent Complexity of the Filters Corresponding to Three Strategies \cite{ACC_2014,Orchard_Thesis} and Our Proposed Scheme.}
\centering
\label{table:compare}
  \scalebox{1.0}{
   \centering
\begin{tabular}{|c|c|c|}
\hline
Prediction Method & Total Equivalent Complexity \\
\hline
   DLM-Based Method \cite{ACC_2014} & $C_A(n_s,n_f ,c_1,c_2,c_3,c_4,N) \; \approx \; $ \\
    & $N(3n_s^2+5n_f^2+6n_f+2n_f n_y+7n_y+3n_s+c_1(n_s+n_f)+c_2(n_s+n_f)+c_3n_s) $  \\
   \hline
  Standard PF-Based Method \cite{Orchard_Thesis} & $C_B(n_s,n_f,c_1,c_3,N) \;\approx \; $ \\
  &  $N(3n_s^2+3n_f^2+6n_sn_f+(1+c_1+c_3)n_s+(1+c_1+c_3)n_f+n_y) $  \\
   \hline
  TTS-EnKF Prediction (this work) & $C_C(n_s,n_f,c_1,N) \; \approx \; $ \\
  & $N(n_s^2+n_f^2+2n_y^2+2n_sn_f+3n_sn_y+3n_fn_y+(9+c_1)n_s+(11+c_1)n_f+9n_y) $ \\
  \hline
\end{tabular}
}
\end{table*}
Given computational complexity of certain common matrix manipulations, as given by \cite{flop}, our goal here is to develop a comprehensive measure and comparison between the complexity of our proposed TTS-EnKF prediction algorithm with other commonly used and well-known particle filtering (PF) prediction schemes \cite{Orchard_2,ACC_2014}.

The EF complexity of the particle filter prediction algorithm with a regularized structure is now provided in Table \ref{table:compare}. This has already been used for prediction purposes in various applications  in \cite{Orchard_Thesis}. We have also included our previously developed prediction algorithm that is based on combination of the particle filters with dynamically linear models (DLM) \cite{ACC_2014}. In Table \ref{table:compare},  $c_1$ denotes the complexity of the random number generation, $c_2$ denotes the complexity of the resampling step of the particle filtering algorithm, $c_3$ denotes the complexity corresponding to the regularization step of the particle filtering algorithm, and $c_4$ denotes the complexity corresponding to the DLM model construction. In Table 1, the EF complexity of the DLM-based prediction method, the standard PF-based prediction method, and the TTS-EnKF prediction method are denoted by $C_A(n_s,n_f ,c_1,c_2,c_3,c_4,N)$, $C_B(n_s,n_f,c_1,c_3,N)$, and $C_C(n_s,n_f,c_1,N)$, respectively. In the above first two methods $N$ represents the number of particles, whereas in the last method $N$ represents the number of ensembles that are selected in the TTS-EnKF approach.

From results  in Table 1, it follows that  PF-based prediction methods requires  a computationally more intensive  cost. This is quantified by the EF complexity (which is a measure of the algorithm time complexity) due to the  resampling ($c_2$) and/or regularization ($c_3$) steps that deal with ordering. These are among the most complex implementation procedures \cite{flop}. In the EF complexity analysis  the operations of order $N$ are considered as dominant that  affect complexity of the entire algorithm.

\section{Development of a health monitoring and prognosis methodology for a
gas turbine engine}\label{V}
The considered application of our proposed two-time scale EnKF method for health monitoring and prognosis of a gas turbine engine is presented in this section. The approach can be used for failure prognosis of the engine, when the system is assumed to be affected by health degradation phenomenon. As demonstrated subsequently our proposed prediction scheme is  capable of tracking the system health parameters that enjoy a slow time dynamics in comparison with the other engine  dynamics  that enjoy a fast time scale behavior. Moreover, the performance of our proposed two-time scale EnKF method is evaluated and investigated under a general degradation scenario in the turbine component due to the erosion phenomenon. The main concept behind  our strategy is to first model the dynamics associated with  the system health parameters and augment them with the gas turbine system states to achieve more accurate estimation as well as prediction results. Therefore, the gas turbine engine   in \cite{nader} is modified in this work to include the dynamical model that is associated with the system health parameters.
\subsection{Overall Model Overview}\label{damage_model_definition}
The formulation for the degradation damage modeling of a gas turbine engine is provided next. In this new methodology the system health parameters, which have a slowly time-varying behavior (due to characteristics of the fault vector), are modeled as state variables with slow dynamics. The most important aspect of this modeling process is that the degradation is assumed to have been initiated  from the beginning  of the engine/turbine operation. This assumption is not very restrictive since a real engine is subject to various types of degradations (such as erosion) from the first initiation of its operation that propagate during future times.

For the class of nonlinear systems that are investigated here (the gas turbine application), the health parameters of the system are denoted by $\theta$ and are considered to be smooth functions of the system states (fast states) and time, i.e., $\theta(x,t)$. The effects of  degradations are modeled by  multiplicative time-varying vector function, $k(t,\epsilon)$, that is known as the fault vector, where $0<\epsilon\ll 1$, is a sufficiently small parameter that quantifies the two time-scale separation characteristics. In other words, the health parameter is represented by
\begin{eqnarray}\label{health}
\vspace{-4mm}
\begin{split}
\theta(x,t)=k(t,\epsilon)\theta _1(x(t)),
\end{split}
\vspace{-4mm}
\end{eqnarray}
where $\theta_1(x(t))$ is a smooth function of $x(t)$. The function $k(t,\epsilon)\in \mathcal{C}^2$ has an asymptotic power series expansion in terms of $\epsilon$ (A function $f(\epsilon)$ has an asymptotic power series expansion if as $\epsilon \rightarrow 0$, $f(\epsilon)\approx \sum _{j=0}^{\infty} f_j \epsilon^j$), \cite{singular_book}, i.e., for $k(t,\epsilon)$ and its first time derivative we have,
\begin{eqnarray}\label{smooth}
\vspace{-4mm}
\begin{split}
k(t,\epsilon)= k^0(t,0)+\epsilon \frac{\partial}{\partial{\epsilon}}k(t,0)+O(\epsilon ^2), k^0(t,0)=0,\\
\dot{k}(t,\epsilon)=\dot{k}^0(t,0)+\epsilon \frac{\partial}{\partial{\epsilon}}{\dot{k}}(t,0)+O(\epsilon ^2),\dot{k}^0(t,0)=0,
\end{split}
\vspace{-4mm}
\end{eqnarray}
where $\dot{k}=\frac{\partial}{\partial {t}}k(t,\epsilon)$. Hence, the  health parameters dynamics augmented with the system states  are given by,
 \begin{eqnarray}\label{dot}
\begin{split}
\dot{\theta}(x,t)=\dot{k}(t,\epsilon)\theta _1(x(t))+k(t,\epsilon)\frac{\partial{\theta _1}}{\partial{x}}\dot{x}(t).
\end{split}
\end{eqnarray}
By considering  series expansions of \eqref{smooth}, the system  equations including the augmented health parameters can be represented in the standard singularly perturbed form by introducing a new time variable $\tau=\epsilon t$, as follows
\begin{eqnarray}\label{sys}
\vspace{-4mm}
\begin{split}
\epsilon \frac{dx}{d\tau}=f(x,\theta,\epsilon,\tau),\\
\frac{d\theta}{d\tau}=g(x,\theta,\epsilon,\tau),
\end{split}
\vspace{-4mm}
\end{eqnarray}
where the time derivatives are taken with respect to $\tau$, $x \in \mathbb{R}^{n_x}$, $\theta \in \mathbb{R}^{n_\theta}$, $f: \mathbb{R}^{n_x}\times \mathbb{R}^{n_{\theta}}\times \mathbb{R}^2\rightarrow \mathbb{R}^{n_x}$, and $g: \mathbb{R}^{n_x}\times \mathbb{R}^{n_{\theta}}\times \mathbb{R}^2\rightarrow \mathbb{R}^{n_{\theta}}$ belong to $\mathcal{C}^2$. In the following simulation scenarios that are conducted the effects of the turbine erosion degradation on the gas turbine system health propagation are investigated. Therefore, the dynamics of the mass flow capacity and efficiency of the turbine are augmented with the system state equations.

The mathematical model of the gas turbine engine that is used in this work is a single spool jet engine that was developed in \cite{nader} and presented in \cite{ACC_khodemoon2}. The four engine states are the combustion chamber pressure and temperature, $P_{\rm{CC}}$ and $T_{\rm{CC}}$, respectively, the spool speed $S$, and the nozzle outlet pressure $P_{\rm{NLT}}$. The continuous-time state space model of the gas turbine is given as follows,
\begin{align}\label{system_eq}
\begin{split}
 & \dot{T}_{\rm{CC}}=\frac{1}{c_{\rm{v}} m_{\rm{cc}}}[(c_{\rm{p}}T_{\rm{C}}{m}_{\rm{C}}+\eta_{\rm{CC}}H_{\rm{u}}{m}_{\rm{f}}
 -c_{\rm{p}}T_{\rm{CC}}\theta_{m_{\rm{T}}})\\
 &
 -c_{\rm{v}}T_{\rm{CC}}({m_{\rm{C}}}+{m}_{\rm{f}}-\theta_{m_{\rm{T}}})],\\
 &\dot{S}=\frac{\eta_{\rm{mech}} \theta_{m_{\rm{T}}}c_{\rm{p}} (T_{\rm{CC}}-T_{\rm{T}}) - {m_{\rm{C}}}c_{\rm{p}} (T_{\rm{C}}-T_{\rm{d}})}{{\rm{J}}S(\frac{\pi}{30})^{2}}, \\
 & \dot{P}_{\rm{CC}}=\frac{P_{\rm{CC}}}{T_{\rm{CC}}}\frac{1}{c_{\rm{v}} m_{\rm{cc}}}[(c_{\rm{p}}T_{\rm{C}}{m_{\rm{C}}}+\eta_{\rm{CC}}H_{\rm{u}}{m}_{\rm{f}}
 -c_{\rm{p}}T_{\rm{CC}}\theta_{m_{\rm{T}}})\\
 &-c_{\rm{v}}T_{\rm{CC}}({m_{\rm{C}}}+{m}_{\rm{f}}-\theta_{m_{\rm{T}}})]
 +\frac{\gamma{\rm{R}}{T_{\rm{CC}}}}{V_{\rm{CC}}}({m_{\rm{C}}}+{m}_{\rm{f}}-\theta_{m_{\rm{T}}}),\\
 & \dot{P}_{\rm{NLT}}=\frac{{T_{\rm{M}}}}{V_{\rm{M}}}(\theta_{m_{\rm{T}}}
 +\frac{\beta}{\beta+1}{m_{\rm{C}}}-{m}_{\rm{Nozzle}}),
\end{split}
\end{align}
The five gas turbine measured outputs are considered to be the compressor temperature ($y_1$), the combustion chamber pressure ($y_2$), the spool speed ($y_3$), the nozzle outlet pressure ($y_4$), and the turbine temperature ($y_5$), namely
$y_{1}=T_{\rm{C}}=T_{\rm{diffuser}}[1+\frac{1}{\eta_{\rm{C}}}
[(\frac{P_{\rm{CC}}}{P_{\rm{diffuzer}}})^{\frac{\gamma-1}{\gamma}}-1]]$,
$y_{2}=P_{\rm{CC}}$, $y_{3}=S$, $y_{4}=P_{\rm{NLT}}$, and
 $y_{5}=T_{\rm{T}}=T_{\rm{CC}}[1-\theta_{\eta_{\rm{T}}}
 (1-(\frac{P_{\rm{NLT}}}{P_{\rm{CC}}})^{\frac{\gamma-1}{\gamma}})]$.
By augmenting the turbine health parameters to the system state equations we obtain
\begin{eqnarray}
\begin{split}
&\dot \theta_{\eta_{\rm{T}}}=\dot{k}_1(t,\epsilon){\eta_{\rm{T}}}(S,P_{\rm{CC}})+k_1(t,\epsilon)(\frac{\partial{\eta_{\rm{T}}}}{\partial{S}}\dot{S}+\frac{\partial{\eta_{\rm{T}}}}{\partial{\beta}}\dot{\beta}),\\
&\dot \theta_{m_{\rm{T}}} =\dot{k}_2(t,\epsilon){m_{\rm{T}}}(S,P_{\rm{CC}})+k_2(t,\epsilon)(\frac{\partial{m_{\rm{T}}}}{\partial{S}}\dot{S}+\frac{m_{\rm{T}}}{\partial{\beta}}\dot{\beta}).
\vspace{-4mm}
\end{split}
\end{eqnarray}
where the physical significance of all the above model parameters is provided in Table \ref{table:2.con_par}, and the functions $k_1(.)$ and $k_2(.)$ model manifestations in the turbine health parameters due to erosion and are considered as polynomial functions with asymptotic series expansion of $\epsilon$. These functions are chosen as $k_1(t,\epsilon)=1-\epsilon t$ and $k_2(t,\epsilon)=1+0.5\epsilon t$, in order to model the erosion degradation as a linear degradation model \cite{naeem_det}. The functions $\eta_{\rm{T}}(S,\beta)$ and $m_{\rm{T}}(S,P_{\rm{CC}})$ are obtained as polynomial functions by utilizing curve-fitting from the turbine performance maps as utilized in \cite{nader}. The details are omitted here due to space limitations.
\begin{table*}
\caption{Model Parameters Description. }
\vspace{-5mm}
\label{table:2.con_par}
\begin{center}
{
\scalebox{1}{
\begin{tabular}{|c||c|c||c|}
\hline
Parameter &Description &Parameter &Description  \\
\hline
\hline
   $c_{\rm{v}}$ &  Specific heat at constant pressure, ${\rm{\frac{J}{kg.K}}}$  & $T_{\rm{T}}$ & Turbine temperature, ${\rm{K}}$    \\
   $c_{\rm{p}}$ &Specific heat at constant volume, ${\rm{\frac{J}{kg.K}}}$ &$T_{\rm{d}}$  &  Intake temperature, ${\rm{K}}$ \\
   $\dot m_{\rm{cc}}$  & Combustion chamber mass flow rate, ${\rm{kg/s}}$   &${\rm{J}}$ & Rotor moment of inertia, ${\rm{kg.m^2}}$   \\
   $T_{\rm{C}}$ & Compressor temperature, ${\rm{K}}$    &${\rm{R}}$  &  Gas constant, ${\rm{\frac{J}{kg.K}}}$     \\
   $H_{\rm{u}}$ &  Fuel specific heat, ${\rm{\frac{J}{kg}}}$  &$\gamma$ & Heat capacity ratio      \\
   $\eta_{\rm{CC}}$ & Combustion chamber efficiency &$V_{\rm{CC}}$  &  Combustion camber Volume, ${\rm{m^3}}$   \\
   $\dot m_{\rm{f}}$  &  Fuel flow, ${\rm{kg/s}}$  &$T_{\rm{M}}$ & Mixer temperature, ${\rm{K}}$     \\
   $\dot m_{\rm{T}}$ & Turbine mass flow rate, ${\rm{kg/s}}$  &$V_{\rm{M}}$  & Mixer volume, ${\rm{m^3}}$   \\
   $\eta_{\rm{T}}$& Turbine efficiency    &$\dot m_{\rm{nozzle}}$ & Nozzle mass flow rate, ${\rm{kg/s}}$   \\
   $\dot m_{\rm{C}}$  & Compressor mass flow rate, ${\rm{kg/s}}$   &$P_{\rm{diffuzer}}$  & Diffuzer pressure, ${\rm{bar}}$   \\
   $\eta_{\rm{C}}$& Compressor efficiency    &$T_{\rm{diffuzer}}$ & Diffuzer temperature, ${\rm{K}}$  \\
      $\eta_{\rm{mech}}$& mechanical efficiency    &$\beta$ & bypass ratio \\
 \hline
\end{tabular}
}}
\end{center}
\vspace{-5mm}
\end{table*}

In order to model the overall gas turbine engine state equations with the  turbine health parameters in the two-time scale framework, it is assumed that $\tau=\epsilon t$, so that one can rewrite the system equation \eqref{system_eq} as
\begin{align}\label{system_eq_tau}
\begin{split}
 &\epsilon \frac{{\rm{d}}{T}_{\rm{CC}}}{{\rm{d}}\tau}=\frac{1}{c_{\rm{v}} m_{\rm{cc}}}[(c_{\rm{p}}T_{\rm{C}}{m}_{\rm{C}}+\eta_{\rm{CC}}H_{\rm{u}}{m}_{\rm{f}}
 -c_{\rm{p}}T_{\rm{CC}}\theta_{m_{\rm{T}}})\\
 &-c_{\rm{v}}T_{\rm{CC}}({m_{\rm{C}}}+{m}_{\rm{f}}-\theta_{m_{\rm{T}}})],\\
 &\epsilon \frac{{\rm{d}}{S}}{{\rm{d}}\tau}=\frac{\eta_{\rm{mech}} \theta_{m_{\rm{T}}}c_{\rm{p}} (T_{\rm{CC}}-T_{\rm{T}}) - {m_{\rm{C}}}c_{\rm{p}} (T_{\rm{C}}-T_{\rm{d}})}{{\rm{J}}S(\frac{\pi}{30})^{2}}, \\
 & \epsilon \frac{{\rm{d}}{P}_{\rm{CC}}}{{\rm{d}}\tau}=\frac{P_{\rm{CC}}}{T_{\rm{CC}}}\frac{1}{c_{\rm{v}} m_{\rm{cc}}}[(c_{\rm{p}}T_{\rm{C}}{m_{\rm{C}}}+\eta_{\rm{CC}}H_{\rm{u}}{m}_{\rm{f}}
 -c_{\rm{p}}T_{\rm{CC}}\theta_{m_{\rm{T}}})\\
 &-c_{\rm{v}}T_{\rm{CC}}({m_{\rm{C}}}+{m}_{\rm{f}}-\theta_{m_{\rm{T}}})]
 +\frac{\gamma{\rm{R}}{T_{\rm{CC}}}}{V_{\rm{CC}}}({m_{\rm{C}}}+{m}_{\rm{f}}-\theta_{m_{\rm{T}}}),\\
 & \epsilon \frac{{\rm{d}}{P}_{\rm{NLT}}}{{\rm{d}}\tau}=\frac{{T_{\rm{M}}}}{V_{\rm{M}}}(\theta_{m_{\rm{T}}}+\frac{\beta}{\beta+1}{m_{\rm{C}}}-{m}_{\rm{Nozzle}}).
\end{split}
\end{align}
Similarly, for the turbine health parameters we have
\begin{eqnarray}
\begin{split}
\frac{{\rm{d}}\theta_{\eta_{\rm{T}}}}{{\rm{d}}\tau} &=-{\eta_{\rm{T}}}(S,\beta)+(1-\tau)(\frac{\partial{\eta_{\rm{T}}}}{\partial{S}}
\frac{{\rm{d}}{S}}{{\rm{d}}\tau}\\
&+\frac{\partial{\eta_{\rm{T}}}}{\partial{\beta}}
(\frac{\partial{\beta}}{\partial{P_{\rm{CC}}}}\frac{{\rm{d}}{P}_{\rm{CC}}}{{\rm{d}}\tau}
+\frac{\partial{\beta}}{\partial{P_{\rm{NLT}}}}\frac{{\rm{d}}{P}_{\rm{NLT}}}{{\rm{d}}\tau})),\\
\frac{{\rm{d}}\theta_{m_{\rm{T}}}}{{\rm{d}}\tau} &={m_{\rm{T}}}(S,\beta)+(1+0.5\tau)(\frac{\partial{m_{\rm{T}}}}{\partial{S}}
\frac{{\rm{d}}{S}}{{\rm{d}}\tau} \\
&+\frac{\partial{m_{\rm{T}}}}{\partial{\beta}}
(\frac{\partial{\beta}}{\partial{P_{\rm{CC}}}}\frac{{\rm{d}}{P}_{\rm{CC}}}{{\rm{d}}\tau}
+\frac{\partial{\beta}}{\partial{P_{\rm{NLT}}}}\frac{{\rm{d}}{P}_{\rm{NLT}}}{{\rm{d}}\tau})).
\end{split}
\end{eqnarray}
The reduced order slow model that is obtained by setting $\epsilon=0$ in \eqref{system_eq_tau}, and substituting $P_{\rm{NLT}}$ from the equation of $y_5$ in $\epsilon \frac{{\rm{d}}{S}}{{\rm{d}}\tau}=0$, yields the following algebraic equations
$T_{\rm{CC}}=\frac{c_pT_{\rm{C}}m_{\rm{C}}+\eta_{\rm{CC}}H_um_f}{c_v(m_c+m_f-\theta_{m_{\rm{T}}})+c_p\theta_{m_{\rm{T}}}},
P_{\rm{CC}}=T_{\rm{CC}}^2 \frac{\gamma R c_v m_{\rm{CC}}}{V_{\rm{CC}}} \frac{(\theta_{m_{\rm{T}}}-m_{\rm{C}}-m_f)}{(c_pT_{\rm{C}}m_{\rm{C}}+\eta_{\rm{CC}}H_u m_f-c_pT_{\rm{CC}}\theta_{m_{\rm{T}}})-c_vT_{\rm{CC}}(m_{\rm{C}}+m_f-\theta_{m_{\rm{T}}})},\\
\beta = \frac{m_{\rm{Nozzle}}-\theta_{m_{\rm{T}}}}{m_{\rm{C}}-m_{\rm{Nozzle}}+\theta_{m_{\rm{T}}}},
P_{\rm{NLT}}=P_{\rm{CC}}(1+\frac{m_{\rm{C}}(T_{\rm{C}}-T_d)}{\eta_{mech}\theta_{m_{\rm{T}}}T_{\rm{CC}}
\theta_{\eta_{\rm{T}}}})^{\frac{\gamma}{\gamma-1}}$.
The terms $\eta_{\rm{T}}$ and $m_{\rm{T}}$ represent polynomial functions of $P_{\rm{CC}}$ and $S$ that are dependent on turbine performance maps. In our simulations we follow a numerical algorithm to compute the derivatives of these maps in terms of $P_{\rm{CC}}$ and $S$.

To discretize the above continuous-time model, the first order approximation of the algorithm that was presented in Remark 1 is used which shows an acceptable result for estimation of both the fast and  slow states of the system with a sampling period of $T_s= 1\;{\rm{msec}}\; (\iota=0.001)$.
\subsection{Simulation Scenarios}
 In the simulation scenarios  considered here the engine is assumed to be subjected to degradation damage due to turbine erosion. This causes a gradual drift in the system health parameters, and as a result the system states. A slowly varying linear degradation model utilized for the turbine health parameters during  $500$ cycles of flight operation that cause a $3\%$ drop in the turbine efficiency and $1.5\%$ increase in its mass flow capacity. The time-scale separation parameter $\epsilon$ is selected as $0.005$ to provide this degradation rate in the turbine.
 \subsection{Erosion Estimation Results}
Our proposed and developed two-time scale filtering methodology introduced in the previous sections is now utilized for estimating the system states as well as the turbine health parameters (as represented by the augmented slow states to the gas turbine state equations). The results of our proposed TTS-EnKF estimation scheme corresponding to the percent of the mean absolute error (MAE$\%$) within an estimation window of $5$ seconds for different number of ensemble members are presented in Table \ref{table:states_Enkf}. The errors  obtained from our method are now compared with the ones that are obtained by using the ``\textit{exact}" EnKF approach (that is when no slow-fast decomposition of the overall states of the system is performed), corresponding to the same number of ensembles. The MAE$\%$ results  are provided in Table \ref{table:states}. It should be pointed out that the ``\textit{exact}" EnKF approach {\underline{does not}} converge, due to numerical ill-conditioning, when the number of ensembles is less than $20$ (N/C in Table \ref{table:states} denotes Not Convergent).
\vspace{-4mm}
\begin{table}[H]
\caption{Estimation MAE$\%$ corresponding to different number of ensembles for the TTS-EnKF method (a) states and (b) measurement outputs.}
\label{table:states_Enkf}
\vspace{-5mm}
\centering
\subtable[]{
\begin{tabular}{|c|c|c|c|c|c|c|}
\hline
State & $N=10 $& $N=50$& $N=100$& $N=200$  \\
\hline
\hline                  
   $P_{\rm{CC}}$ &0.7481    &0.7440    &0.6532    &0.6510\\
  $N$  &0.1185    &0.0806    &0.0515    &0.0495    \\
  $T_{\rm{CC}}$& 0.1220    &0.0668    &0.0613    &0.0611  \\
   $P_{\rm{NLT}}$ & 1.1822    &1.1774    &0.9521    &0.9213 \\
   $ \theta_{\eta_{\rm{T}}}$ &  0.6831    &0.5938    &0.4281    &0.4210  \\
   $\theta_{m_{\rm{T}}}$ &0.0614    &0.0342    &0.0322   & 0.0341   \\
\hline
\end{tabular}
}
\centering
\subtable[]{
\begin{tabular}{|c|c|c|c|c|c|c|c|}
\hline
Output & $N=10$& $N=50$& $N=100$& $N=200$  \\
\hline
\hline                  
   $T_{\rm{C}}$  &0.4118    &0.3013    &0.2451    &0.2510   \\
  $P_{\rm{C}}$  &  1.5231    &1.4867    &1.3047    &1.3045     \\
  $N$&   1.1148    &0.0806    &0.0655    &0.06122   \\
   $T_{\rm{T}}$ &  0.3147    &0.2338    &0.2001   & 0.2170    \\
   $P_{\rm{T}}$ &  2.6250   & 2.6287    &2.2830    &2.3030  \\
\hline
\end{tabular}
}
\end{table}
It should be noted that the covariance matrix in the \textit{exact} EnKF method is dependent on the time-scale separation parameter $\epsilon$. This as a matter of fact  can cause non-singularity of the covariance matrix under certain scenarios, and consequently  divergence of the Kalman filtering algorithm due to ill-conditioning and ill-posedness of the estimation problem (in computing the Kalman gain). This limitation of the exact EnKF is more pronounced  for smaller values of $\epsilon$ and/or lower number of ensembles in the implemented EnKF.

A comparison between the TTS-EnKF and the exact EnKF estimation results shows that although the exact method is \textit{not} capable of performing the system state estimation for \textit{lower} number of ensembles, the TTS-EnKF approach is still capable of performing the estimation objective with {\underline{a fewer number}} of ensembles, and consequently it can yield a less computationally costly implementation strategy.
\begin{table}[H]
\caption{Estimation MAE$\%$ corresponding to different number of ensembles for the exact EnKF method (a) states and (b) measurement outputs (N/C denotes not convergent).}
\label{table:states}
\vspace{-5mm}
\centering
\subtable[]{
\begin{tabular}{|c|c|c|c|c|c|c|}
\hline
State & $N=10 $& $N=50$& $N=100$& $N=200$  \\
\hline
\hline                  
   $P_{\rm{CC}}$ &N/C    &0.3355    &0.3022    &0.3020\\
  $N$  &N/C    &0.0504    &0.0492    &0.0497    \\
  $T_{\rm{CC}}$& N/C    &0.0714    &0.0661    &0.0670  \\
   $P_{\rm{NLT}}$ & N/C    &0.2254    &0.2142    &0.2145 \\
   $ \theta_{\eta_{\rm{T}}}$ &  N/C   &0.3021    &0.2815    &0.2781  \\
   $\theta_{m_{\rm{T}}}$ &N/C   &0.0746    &0.0526   & 0.0532   \\
\hline
\end{tabular}
}
\centering
\subtable[]{
\begin{tabular}{|c|c|c|c|c|c|c|c|}
\hline
Output & $N=10$& $N=50$& $N=100$& $N=200$  \\
\hline
\hline                  
   $T_{\rm{C}}$  &N/C    &0.1589    &0.1322    &0.1323   \\
  $P_{\rm{C}}$  &  N/C    &1.1821    &1.1620    &1.1400     \\
  $N$&   N/C    &0.0504    &0.0454    &0.0427   \\
   $T_{\rm{T}}$ &  N/C    &0.1353    &0.1132   & 0.1151    \\
   $P_{\rm{T}}$ &  N/C   & 2.3484    &2.2550    &2.2260  \\
\hline
\end{tabular}
\vspace{-9mm}
}
\end{table}
\begin{table}[H]
\caption{Estimation MAE$\%$ corresponding to different values of $\epsilon$ for the TTS-EnKF method (a) states and (b) measurement outputs.}
\label{table:ep.tts}
\vspace{-5mm}
\centering
\subtable[]{
\begin{tabular}{|c|c|c|c|c|c|c|}
\hline
State & $\epsilon=0.005 $& $\epsilon=0.003$& $\epsilon=0.001$& $\epsilon=0.0001$  \\
\hline
\hline                  
   $P_{\rm{CC}}$ &0.6532    &0.6481    &0.6320    &0.6505\\
  $N$  &0.0515    &0.05000    &0.05325    &0.05260    \\
  $T_{\rm{CC}}$& 0.0613    &0.0608    &0.0615    &0.0611  \\
   $P_{\rm{NLT}}$ & 0.9521    &0.9484    &0.9511    &0.9491 \\
   $ \theta_{\eta_{\rm{T}}}$ &  0.4281    &0.4312    &0.4255    &0.4380  \\
   $\theta_{m_{\rm{T}}}$ &0.0322    &0.0356    &0.0327   & 0.0351   \\
\hline
\end{tabular}
}
\centering
\subtable[]{
\begin{tabular}{|c|c|c|c|c|c|c|c|}
\hline
Output & $\epsilon=0.005 $& $\epsilon=0.003$& $\epsilon=0.001$& $\epsilon=0.0001$  \\
\hline
\hline                  
   $T_{\rm{C}}$  &0.2451    &0.2266    &0.2219    &0.2205   \\
  $P_{\rm{C}}$  &  1.3047    &1.3164    &1.3083    &1.3085     \\
  $N$&   0.0655    &0.0602    &0.0637    &0.0652   \\
   $T_{\rm{T}}$ &  0.2001    &0.2109    &0.2012   & 0.2149    \\
   $P_{\rm{T}}$ &  2.2830   & 2.2583    &2.2530    &2.3072  \\
\hline
\end{tabular}
\vspace{-3mm}
}
\end{table}
\begin{table}[H]
\caption{Estimation MAE$\%$ corresponding to different values of $\epsilon$ for the exact EnKF method (a) states and (b) measurement outputs (N/C denotes not convergent).}
\label{table:ep.ex}
\vspace{-5mm}
\centering
\subtable[]{
\begin{tabular}{|c|c|c|c|c|c|c|}
\hline
State & $\epsilon=0.005 $& $\epsilon=0.003$& $\epsilon=0.001$& $\epsilon=0.0001$  \\
\hline
\hline                  
   $P_{\rm{CC}}$ &0.3022    &0.6255    &N/C    &N/C\\
  $N$  &0.0492    &0.0651    &N/C    &N/C    \\
  $T_{\rm{CC}}$& 0.0661    &0.1200    &N/C    &N/C  \\
   $P_{\rm{NLT}}$ & 0.2142    &0.3541    &N/C    &N/C \\
   $ \theta_{\eta_{\rm{T}}}$ &  0.2815   &0.4537    &N/C    &N/C  \\
   $\theta_{m_{\rm{T}}}$ &0.0526   &0.1070    &N/C   & N/C   \\
\hline
\end{tabular}
}
\centering
\subtable[]{
\begin{tabular}{|c|c|c|c|c|c|c|c|}
\hline
Output & $\epsilon=0.005 $& $\epsilon=0.003$& $\epsilon=0.001$& $\epsilon=0.0001$  \\
\hline
\hline                  
   $T_{\rm{C}}$  &0.1322    &0.2220    &N/C    &N/C   \\
  $P_{\rm{C}}$  &  1.1620    &1.2811    &N/C    &N/C     \\
  $N$&   0.0454    &0.0567    &N/C    &N/C   \\
   $T_{\rm{T}}$ & 0.1132    &0.2112    &N/C   & N/C    \\
   $P_{\rm{T}}$ &  2.2550   & 2.6372    &N/C    &N/C  \\
\hline
\end{tabular}
}
\vspace{-3mm}
\end{table}
The results provided in Table \ref{table:states_Enkf} indicate that by increasing the number of ensembles to more than $N=100$ \textit{does not} necessarily result in a more accurate estimation performance. The best estimation results  achieved are for $100$ ensembles with the maximum percentage of the mean absolute error (MAE$\%$) of $0.95\%$ for the state estimation obtained for the nozzle pressure, and the MAE$\%$ of $2.28\%$ for the output estimation obtained for the turbine pressure. However, due to approximations that we have made to obtain the algebraic equations as described  in Subsection \ref{damage_model_definition} for the turbine health parameters, in this specific scenario with $\epsilon=0.005$, in almost all cases, the exact EnKF method results in a lower MAE$\%$ for both the state and output estimations. Moreover, the discrepancy between the TTS-EnKF and the exact EnKF approaches corresponding to the output estimation problem is lower than that of the state  estimation problem.

To illustrate the effects of $\epsilon$ on the performance of both exact EnKF and TTS-EnKF methods, the degradation scenario is repeated with different values of  $\epsilon$ with $N=100$. These results are provided in Tables \ref{table:ep.tts} and \ref{table:ep.ex}. The summarized results in these two tables show that the estimation accuracy of the TTS-EnKF is not affected by $\epsilon$, whereas the exact EnKF estimation accuracy is highly dependent on $\epsilon$ where for $\epsilon \le 0.001$ the algorithm becomes ill-conditioned and \underline{cannot} converge.
\subsection{Erosion Prediction Results}
In this case study scenario, our prediction strategy developed based on the two-time scale EnKF method is utilized to predict the propagation of the system states (fast states) and the turbine health parameters (slow states) when the degradation due to the erosion has affected the system during its entire operating horizon (that is, $500$ cycles of flight).

For the prediction case study, $N=100$ is selected for both  TTS-EnKF and  exact EnKF schemes. Our prediction case study also includes comparisons with  classical PF-based prediction method \cite{Orchard_2} (as provided in Subsection \ref{Complexity}) by using $100$ particles in order to evaluate  the execution times of all the three methods as a measure of the EF complexity as described in Subsection \ref{Complexity}.

The prediction horizon is  extended from  $100$  to $500$ steps-ahead and  MAE$\%$ results for first and  last prediction windows are provided in Tables \ref{table:states_100} and \ref{table:states_500}, respectively. From these results  one can conclude that the PF-based prediction algorithm with $100$ particles {\underline{does not}} show an acceptable prediction performance.

In other words, beyond the $100$ steps-ahead prediction horizon the MAE$\%$ increases drastically corresponding to both  state and output prediction results. On the other hand, as the prediction horizon is extended, the MAE$\%$ also increases corresponding to the prediction results associated with both  exact EnKF and  TTS-EnKF approaches. However, the TTS-EnKF scheme yields a more robust prediction accuracy results as compared to the exact EnKF method. For example, the maximum $100$ steps-ahead MAE$\%$ for $\theta_{\eta_{\rm{T}}}$ prediction using the TTS-EnKF method is $0.34\%$,  whereas it is around $0.42\%$ for the exact EnKF scheme. We emphasize here again that for this specific scenario with $\epsilon=0.005$ the exact EnKF method \underline{does not} converge.

Finally, the execution time (or equivalently the EF) associated with one iteration of each scheme is  provided in Table \ref{table:Complexity}. The results show a large difference between the PF-based prediction scheme execution time and that associated with and compared with the EnKF-based approaches.
\begin{table}[H]
\caption{100-step ahead prediction MAE$\%$ corresponding to three different methods (a) states and (b) measurement outputs.}
\label{table:states_100}
\vspace{-5mm}
\centering
\subtable[]{
\begin{tabular}{|c|c|c|c|}
\hline
State & TTS-EnKF& Exact EnKF& PF-Based Method  \\
\hline
\hline                  
   $P_{\rm{CC}}$ &0.2118    &0.2843    &0.4149\\
  $N$  &0.0474    &0.0816    &0.1017\\
  $T_{\rm{CC}}$&0.1220    &0.1437    &0.1653\\
   $P_{\rm{NLT}}$ &0.2854    &0.3778    &0.6373\\
   $ \theta_{\eta_{\rm{T}}}$ &  0.3439    &0.4283    &0.5030     \\
   $\theta_{m_{\rm{T}}}$ &0.0087    &0.0099    &0.0109     \\
\hline
\end{tabular}
}
\centering
\subtable[]{
\begin{tabular}{|c|c|c|c|}
\hline
Output & TTS-EnKF& Exact EnKF& PF-Based Method  \\
\hline
\hline                  
   $T_{\rm{C}}$  &0.1052    &0.1232    &0.1895      \\
  $P_{\rm{C}}$  &  1.3338    &1.3285    &1.3801      \\
  $N$&   0.0474    &0.0816    &0.1017       \\
   $T_{\rm{T}}$ &  0.1989    &0.2090    &0.1937      \\
   $P_{\rm{T}}$ &  1.8963   & 1.8666    &1.9765  \\
\hline
\end{tabular}
}
\vspace{-3mm}
\end{table}
\begin{table}[H]
\caption{500-step ahead prediction MAE$\%$ corresponding to three different methods (a) states and (b) measurement outputs.}
\label{table:states_500}
\vspace{-5mm}
\centering
\subtable[]{
\begin{tabular}{|c|c|c|c|}
\hline
State & TTS-EnKF& Exact EnKF& PF-Based Method  \\
\hline
\hline                  
   $P_{\rm{CC}}$ &1.0542    &1.2994    &3.5630\\
  $N$  &0.5168    &0.6374    &1.7717\\
  $T_{\rm{CC}}$&0.5700    &0.7287    &1.9753\\
   $P_{\rm{NLT}}$ &1.2063    &1.5131    &4.0037\\
   $ \theta_{\eta_{\rm{T}}}$ &  1.8358    &2.1622    &6.4120     \\
   $\theta_{m_{\rm{T}}}$ &0.0287    &0.0342    &0.1037     \\
\hline
\end{tabular}
}
\centering
\subtable[]{
\begin{tabular}{|c|c|c|c|}
\hline
Output & TTS-EnKF& Exact EnKF& PF-Based Method  \\
\hline
\hline                  
   $T_{\rm{C}}$  &0.3993    &0.4859    &1.4007      \\
  $P_{\rm{C}}$  &  1.6270    &1.7990    &3.8850      \\
  $N$&   0.5168    &0.6374    &1.7717       \\
   $T_{\rm{T}}$ &  1.1358    &1.3814    &3.9448      \\
   $P_{\rm{T}}$ &  2.2675   & 2.3229   &4.1315  \\
\hline
\end{tabular}
}
\vspace{-3mm}
\end{table}

\begin{table}[H]
\caption{Time Complexity Analysis (and equivalently EF analysis) for the TTS-EnKF, Exact EnKF and the PF-Based Prediction Methods in seconds corresponding to one iteration of each scheme.}
\centering
\vspace{-2mm}
\label{table:Complexity}
  \scalebox{1}{
   \centering
\begin{tabular}{|c|c|c|c|}
\hline
\multirow{1}{*}{Method} & \multicolumn{1}{c|}{Best Case} & \multicolumn{1}{c|}{Average Case}& \multicolumn{1}{c|}{Worst Case}  \\ 
\hline
   TTS-EnKF &     1.1676 &  1.3112 &  2.9235    \\
   Exact EnKF &    0.8898 &  0.9310 &  1.0020    \\
   PF-based &    24.4211 &  33.5490 &  64.0247    \\
   \hline
\end{tabular}
}
\vspace{-3mm}
\end{table}
\section{Conclusion}\label{VI}
In this paper, first a novel two-time scale estimation filter is developed and designed for a nonlinear system based on an ensemble Kalman filtering (En-KF) approach to estimate its fast and slow states. One of the main application of our proposed estimation strategy is in investigating the health monitoring and damage tracking problems of a nonlinear system. Based on our developed estimation algorithm, a two-time scale prediction methodology is then proposed to predict the long-term behavior of the system states. Our proposed estimation and prediction methodologies were applied to a gas turbine engine system to illustrate and validate our results when the engine system is affected by a gradual degradation damage due to erosion. The resulting estimation and prediction observations indicate an acceptable performance of our methods and confirm that our strategy is quite suitable for further investigation in the domain of health and condition-based monitoring research.



\begin{thebibliography}{10}

\bibitem{luo2008}
J.~Luo, K.~R. Pattipati, L.~Qiao, and S.~Chigusa, ``Model-based prognostic
  techniques applied to a suspension system,'' {\em IEEE Transactions on
  Systems, Man and Cybernetics, Part A: Systems and Humans}, vol.~38, no.~5,
  pp.~1156--1168, 2008.

\bibitem{model1}
J.~Cusumano and A.~Chatterjee, ``A dynamical systems approach to damage
  evolution tracking, part 1: description and experimental application,'' {\em
  Journal of Vibration and Acoustics}, vol.~124, no.~2, pp.~250--257, 2002.

\bibitem{gillijns2006}
S.~Gillijns, O.~B. Mendoza, J.~Chandrasekar, B.~De~Moor, D.~Bernstein, and
  A.~Ridley, ``What is the ensemble kalman filter and how well does it work?,''
  in {\em American Control Conference}, pp.~6--pp, 2006.

\bibitem{daum1986}
F.~E. Daum, ``Exact finite-dimensional nonlinear filters,'' {\em IEEE
  Transactions on Automatic Control}, vol.~31, no.~7, pp.~616--622, 1986.

\bibitem{krener1996}
A.~Krener and A.~Duarte, ``A hybrid computational approach to nonlinear
  estimation,'' in {\em Proceedings of the 35th IEEE Conference on Decision and
  Control,}, vol.~2, pp.~1815--1819, IEEE, 1996.

\bibitem{smith2013}
A.~Smith, A.~Doucet, N.~de~Freitas, and N.~Gordon, {\em Sequential Monte Carlo
  methods in practice}.
\newblock Springer Science \& Business Media, 2013.

\bibitem{gelb1974}
A.~Gelb, {\em Applied optimal estimation}.
\newblock MIT press, 1974.

\bibitem{evensen2003}
G.~Evensen, ``The ensemble kalman filter: Theoretical formulation and practical
  implementation,'' {\em Ocean dynamics}, vol.~53, no.~4, pp.~343--367, 2003.

\bibitem{gillijns2007}
S.~Gillijns and B.~De~Moor, ``Model error estimation in ensemble data
  assimilation,'' {\em Nonlinear Processes in Geophysics}, vol.~14, no.~1,
  pp.~59--71, 2007.

\bibitem{triantafyllou2013}
G.~Triantafyllou, I.~Hoteit, X.~Luo, K.~Tsiaras, and G.~Petihakis, ``Assessing
  a robust ensemble-based kalman filter for efficient ecosystem data
  assimilation of the cretan sea,'' {\em Journal of Marine Systems}, vol.~125,
  pp.~90--100, 2013.

\bibitem{ott2004}
E.~Ott, B.~R. Hunt, I.~Szunyogh, A.~V. Zimin, E.~J. Kostelich, M.~Corazza,
  E.~Kalnay, D.~Patil, and J.~A. Yorke, ``A local ensemble kalman filter for
  atmospheric data assimilation,'' {\em Tellus A}, vol.~56, no.~5,
  pp.~415--428, 2004.

\bibitem{evensen2009}
G.~Evensen, ``The ensemble kalman filter for combined state and parameter
  estimation,'' {\em IEEE Control Systems}, vol.~29, no.~3, pp.~83--104, 2009.

\bibitem{anderson1999}
J.~L. Anderson and S.~L. Anderson, ``A monte carlo implementation of the
  nonlinear filtering problem to produce ensemble assimilations and
  forecasts,'' {\em Monthly Weather Review}, vol.~127, no.~12, pp.~2741--2758,
  1999.

\bibitem{sakov2008}
P.~Sakov and P.~R. Oke, ``A deterministic formulation of the ensemble kalman
  filter: an alternative to ensemble square root filters,'' {\em Tellus A},
  vol.~60, no.~2, pp.~361--371, 2008.

\bibitem{kwiatkowski2015}
E.~Kwiatkowski and J.~Mandel, ``Convergence of the square root ensemble kalman
  filter in the large ensemble limit,'' {\em SIAM/ASA Journal on Uncertainty
  Quantification}, vol.~3, no.~1, pp.~1--17, 2015.

\bibitem{lorentzen2011}
R.~J. Lorentzen and G.~N{\ae}vdal, ``An iterative ensemble kalman filter,''
  {\em IEEE Transactions on Automatic Control}, vol.~56, no.~8, pp.~1990--1995,
  2011.

\bibitem{khalil1996nonlinear}
H.~K. Khalil and J.~Grizzle, {\em Nonlinear systems}, vol.~3.
\newblock Prentice hall New Jersey, 1996.

\bibitem{rauch}
H.~Rauch, ``Order reduction in estimation with singular perturbation,'' in {\em
  4 th Symposium on Nonlinear Estimation Theory and Its Applications},
  pp.~231--241, 1974.

\bibitem{gajic}
Z.~Gajic and M.~Lim, ``A new filtering method for linear singularly perturbed
  systems,'' {\em IEEE Transactions on Automatic Control}, vol.~39, no.~9,
  pp.~1952--1955, 1994.

\bibitem{shen}
X.~Shen, M.~Rao, and Y.~Ying, ``Decomposition method for solving {K}alman
  filter gains in singularly perturbed systems,'' {\em Optimal Control
  Applications and Methods}, vol.~14, no.~1, pp.~67--73, 1993.

\bibitem{naidu_1}
D.~S. Naidu and A.~K. Rao, {\em Singular perturbation analysis of discrete
  control systems}, vol.~1154.
\newblock Springer-Verlag Berlin, 1985.

\bibitem{udem}
M.~Aliyu and E.~Boukas, ``{H}$_\infty$-filtering for singularly perturbed
  nonlinear systems,'' {\em International Journal of Robust and Nonlinear
  Control}, vol.~21, no.~2, pp.~218--236, 2011.

\bibitem{pf}
J.~H. Park, H.~C. Yeong, and N.~S. Namachchivaya, ``Particle filters in a
  multiscale environment: homogenized hybrid particle filter,'' {\em Journal of
  Applied Mechanics}, vol.~78, no.~6, p.~061001, 2011.

\bibitem{socha2000exponential}
L.~Socha, ``Exponential stability of singularly perturbed stochastic systems,''
  {\em IEEE Transactions on Automatic Control}, vol.~45, no.~3, pp.~576--580,
  2000.

\bibitem{barbot1991using}
J.~P. Barbot and N.~Pantalos, ``Using symbolic calculus for singularly
  perturbed nonlinear systems,'' in {\em Algebraic Computing in Control},
  pp.~40--49, Springer, 1991.

\bibitem{khalil}
P.~Kokotovic, H.~Khali, and J.~O'reilly, {\em Singular perturbation methods in
  control: analysis and design}, vol.~25.
\newblock Society for Industrial Mathematics, 1987.

\bibitem{spong1987integral}
M.~Spong, K.~Khorasani, and P.~V. Kokotovic, ``An integral manifold approach to
  the feedback control of flexible joint robots,'' {\em IEEE Journal on
  Robotics and Automation}, vol.~3, no.~4, pp.~291--300, 1987.

\bibitem{barbot1996analysis}
J.~Barbot, M.~Djemai, S.~Monaco, and D.~Normand-Cyrot, ``Analysis and control
  of nonlinear singularly perturbed systems under sampling,'' {\em Control and
  Dynamic Systems}, vol.~79, pp.~203--246, 1996.

\bibitem{barbot1991}
J.~P. Barbot, S.~Monaco, D.~Normand-Cyrot, and N.~Pantalos, ``Discretization
  schemes for nonlinear singularly perturbed systems,'' in {\em Proceedings of
  the 30th IEEE Conference on Decision and Control}, pp.~443--448, 1991.

\bibitem{teixeira2008gain}
B.~O.~S. Teixeira, J.~Chandrasekar, H.~J. Palanthandalam-Madapusi, L.~A.~B.
  T{\^o}rres, L.~A. Aguirre, and D.~S. Bernstein, ``Gain-constrained kalman
  filtering for linear and nonlinear systems,'' {\em IEEE Transactions on
  Signal Processing}, vol.~56, no.~9, pp.~4113--4123, 2008.

\bibitem{mandel2011}
J.~Mandel, L.~Cobb, and J.~D. Beezley, ``On the convergence of the ensemble
  kalman filter,'' {\em Applications of Mathematics}, vol.~56, no.~6,
  pp.~533--541, 2011.

\bibitem{inequalities}
G.~Hardy, J.~Littlewood, and G.~Polya, ``Inequalities. reprint of the 1952
  edition. cambridge mathematical library,'' 1988.

\bibitem{ACC_2014}
N.~Daroogheh, N.~Meskin, and K.~Khorasani, ``A novel particle filter parameter
  prediction scheme for failure prognosis,'' in {\em American Control
  Conference, 2014}, pp.~1735--1742, 2014.

\bibitem{Orchard_Thesis}
M.~E. Orchard, {\em A Particle Filtering-based Framework for On-line Fault
  Diagnosis and Failure Prognosis}.
\newblock PhD thesis, Georgia Institute of Technology, 2006.

\bibitem{flop}
R.~Karlsson, T.~Sch{\"o}n, and F.~Gustafsson, ``Complexity analysis of the
  marginalized particle filter,'' {\em IEEE Transactions on Singnal
  Processing}, vol.~53, no.~11, pp.~4408--4411, 2005.

\bibitem{Orchard_2}
A.~Doucet, N.~De~Freitas, and N.~Gordon, {\em Sequential {M}onte {C}arlo
  methods in practice}.
\newblock Springer Verlag, 2001.

\bibitem{nader}
E.~Naderi, N.~Meskin, and K.~Khorasani, ``Nonlinear fault diagnosis of jet
  engines by using a multiple model-based approach,'' {\em Journal of
  Engineering for Gas Turbines and Power}, vol.~134, no.~1, p.~011602, 2012.

\bibitem{singular_book}
J.~R. O'Malley, ``Singular perturbation methods for ordinary differential
  equations,'' 1990.

\bibitem{ACC_khodemoon2}
N.~Daroogheh, N.~Meskin, and K.~Khorasani, ``A novel particle filter parameter
  prediction scheme for failure prognosis,'' in {\em Proceedings of the
  American Control Conference}, pp.~1735--1742, 2014.

\bibitem{naeem_det}
M.~Naeem, R.~Singh, and D.~Probert, ``Consequences of aero-engine
  deteriorations for military aircraft,'' {\em Applied Energy}, vol.~70, 2001.

\end{thebibliography}

\end{document}